\documentclass[fleqn,10pt]{wlscirep}
\usepackage{amssymb,amsmath,enumerate}
\usepackage{multirow}
\usepackage{bm}
\usepackage{epsfig}
\usepackage{multicol}        
\usepackage{url}
\usepackage{authblk}

\title{Characterization of catastrophic instabilities: Market crashes as paradigm}

\author[1,*]{Anirban Chakraborti}
\author[1]{Kiran Sharma}
\author[2]{Hirdesh K. Pharasi}
\author[3]{Sourish Das}
\author[2,4]{Rakesh Chatterjee}
\author[2,5]{Thomas H. Seligman}

\affil[1]{School of Computational and Integrative Sciences, Jawaharlal Nehru University, New Delhi-110067, India}
\affil[2]{Instituto de Ciencias F\'{i}sicas, Universidad Nacional Aut\'{o}noma de M\'{e}xico, Cuernavaca-62210, M\'{e}xico}
\affil[3]{Chennai Mathematical Institute, Chennai-603103, India}
\affil[4]{School of Mechanical Engineering,
Tel Aviv University, Israel}
\affil[5]{Centro Internacional de Ciencias, Cuernavaca-62210, M\'{e}xico}

\affil[*]{anirban@jnu.ac.in}

\begin{abstract}

Catastrophic events, though rare, do occur and when they occur, they have devastating effects. It is, therefore, of utmost importance to understand the complexity of the underlying dynamics and signatures of catastrophic events, such as market crashes. For deeper understanding, we choose the US and Japanese markets from 1985 onward, and study the evolution of the cross-correlation structures of stock return matrices and their eigenspectra over different short time-intervals or “epochs”. A slight non-linear distortion is applied to the correlation matrix computed for any epoch, leading to the {\it emerging spectrum} of eigenvalues. The statistical properties of the emerging spectrum display: (i) the shape of the emerging spectrum reflects the market instability, (ii) the smallest eigenvalue may be able to statistically distinguish the nature of a market turbulence or crisis -- internal instability or external shock, and (iii) the time-lagged smallest eigenvalue has a statistically significant correlation with the mean market cross-correlation. The smallest eigenvalue seems to indicate that the financial market has become more turbulent in a similar way as the mean does. Yet we show features of the smallest eigenvalue of the emerging spectrum that distinguish different types of market instabilities related to internal or external causes. Based on the paradigmatic character of financial time series for other complex systems, the capacity of the emerging spectrum to understand the nature of instability may be a new feature, which can be broadly applied.
\end{abstract}

\begin{document}

\flushbottom
\maketitle

\thispagestyle{empty}

\section*{Introduction}

A stock market is a fascinating example of a complex system\cite{Vemuri1978,Gellmann1995,Yaneer2002}, where the coherent collective behavior of the economic agents and their repeated nonlinear interactions, often lead to rich structures of correlations and time-dependencies \cite{Mantegna_Stanley_book,bouchaud2003,sinha2010}. The movements in the market prices are often influenced by news or external shocks, which can result in the unforeseen and rapid drop in the prices of a large section of the stock market, labeled as a market crash! On the contrary, the widespread existence of bubbles in financial markets and extreme movements of return series often result from the unstable relationship between macroeconomic fundamentals of the economy
and the asset prices \cite{Shiller_AER_81}.
Since the societal impact of an extreme event like a market crash can be catastrophic \cite{Sornette_04,Buchanan2000}, 
the understanding of such events \cite{Sorkin_09}, the assessment of the associated risks \cite{Acemoglu_15}, and possible prediction of these events have drawn attention from all quarters: governments, industry participants and academia.
Recent research has emphasized the roles played by bounded rationality as important causal factors for the observed disconnect between volatility of asset returns
and movements of the underlying fundamentals, or the `excess volatility puzzle' \cite{Sornette_04,Gabaix_12}. A recent paper  \cite{Kiran2017} presented the alternative view that the co-movements in financial assets are anchored to the corresponding macroeconomic fundamentals -- the nominal returns from individual assets might drift far from what can be predicted using expected cash-flow, while the joint evolution of the co-movements of returns are still related to aggregate size variables like market capitalization, revenue or number of employees. There already exist papers related to economic fluctuations using network theory \cite{Acemoglu_12,Acemoglu_net_macro_16}, but there is a lot more to understand about movements and fluctuations of markets. 

It is widely accepted that complex financial markets are not amenable to mathematically analytic description nor to computational reduction -- financial markets inherently unpredictable and the only way
to study their evolution is perhaps to let them evolve in time!
However, the detailed evolution of a complex system may not have much significance; it may be more interesting to study certain phases of the evolution, like rare events. A forecasting or prediction algorithm may be required for answering important questions related to rare events -- predicting the occurrence of seismic waves or temperature rise, forecasting crashes or bubbles, etc. Such extreme events often reveal underlying dynamical processes and thus provide ground for a better scientific understanding of complex systems like stock markets, fractures or earthquakes \cite{Bikas,Seligman}. 
Theories and concepts from statistical physics have often proven to be of much use in studying and understanding the collective behavior in complex financial markets \cite{sinha2010}. Tools from random matrix theory (RMT) \cite{Mehta2004}, self-organized criticality \cite{Bak1996}, networks \cite{Barabasi2002}, etc. have been used extensively to model such complex systems \cite{Goldenfeld1999}.

We present a study of the time evolution of the cross-correlation structures of return matrices for $N$ stocks, and the eigenspectra over different time-epochs, as traditionally analyzed using tools of RMT. 
Correlation matrices have been used as a standard tool in the analysis of the time evolution of complex systems, and particularly in financial markets \cite{OnnelaI:03,Munnix2012}. For this type of analysis, one assumes stationarity of the underlying time series. As this assumption manifestly fails for longer time series, it is often useful to break the long time series of length $T$, into $n$ time-epochs of size $M$ (such that $T/M=n$). The assumption of stationarity  improves for the shorter time-epochs used. However, if there are $N$ return time series such that $N > M$, this implies an analysis of highly singular correlation matrices with $N-M+1$ zero eigenvalues, which lead to poor statistics. This problem can be avoided by using the non-linear ``power map'', which was introduced to reduce noise and break the degeneracy of the zero eigenvalues \cite{Guhr2003,Guhr2010,Vinayak2013}.  In this paper, we have  used a small non-linear distortion (foregoing the noise reduction) for each cross-correlation matrix computed within an epoch, giving rise to the ``emerging spectrum'' of eigenvalues well separated from the original non-zero ones, which can be then used as a ``signal'' \cite{Vinayak2013}. We then study the statistical properties of the emerging spectrum and show for the first time that: (i) the shape of the emerging spectrum reflects the market turbulence, and (ii) the smallest eigenvalue of the emerging spectrum may be able to statistically distinguish the principal nature of a market turbulence or crisis -- internal reaction or external shock. Further, a linear regression model of the mean market cross-correlation as a function of the time-lagged smallest eigenvalue indicates, that the two variables have statistically significant correlation. For the USA, from 2001 onward, the smallest eigenvalue seems to indicate that the financial market has become more turbulent. Similarly for Japan, the nature of the market has changed from 1990 onward. For this, we use the GARCH($ p,q $) model \cite{Tsay} to estimate the volatility of a time series.

\section*{Methodology and Results}
We take the different time series of the logarithmic returns $r_i$, of the stocks in the USA and JPN markets (see Data description), and construct the equal-time cross-correlation matrix, with elements: $ C_{ij}(t) = (\langle r_i r_j \rangle - \langle r_i \rangle \langle r_j \rangle)/\sigma_i\sigma_j$, where $i,j=1, \dots, N$ and $t$ indicates the end date of the time-epoch of size $M$. 
The correlations are computed over time-epochs of $M=20$ days for the entire return series $T=8067$ days for USA, and $T=7997$ days for JPN. We study the evolution of the cross-correlation structures of return matrices $\boldsymbol{C}(t)$ and the eigenspectra over different overlapping time-epochs (shifted by $\Delta t=1$ day). 

\begin{figure}
\centering
\includegraphics[width=0.95\linewidth]{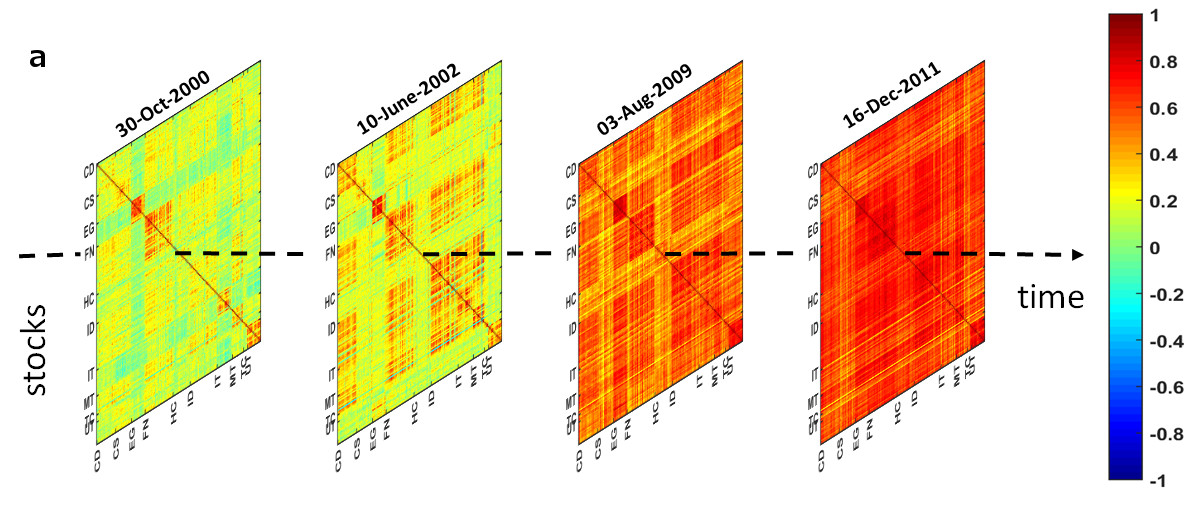}\\
\includegraphics[width=0.33\linewidth]{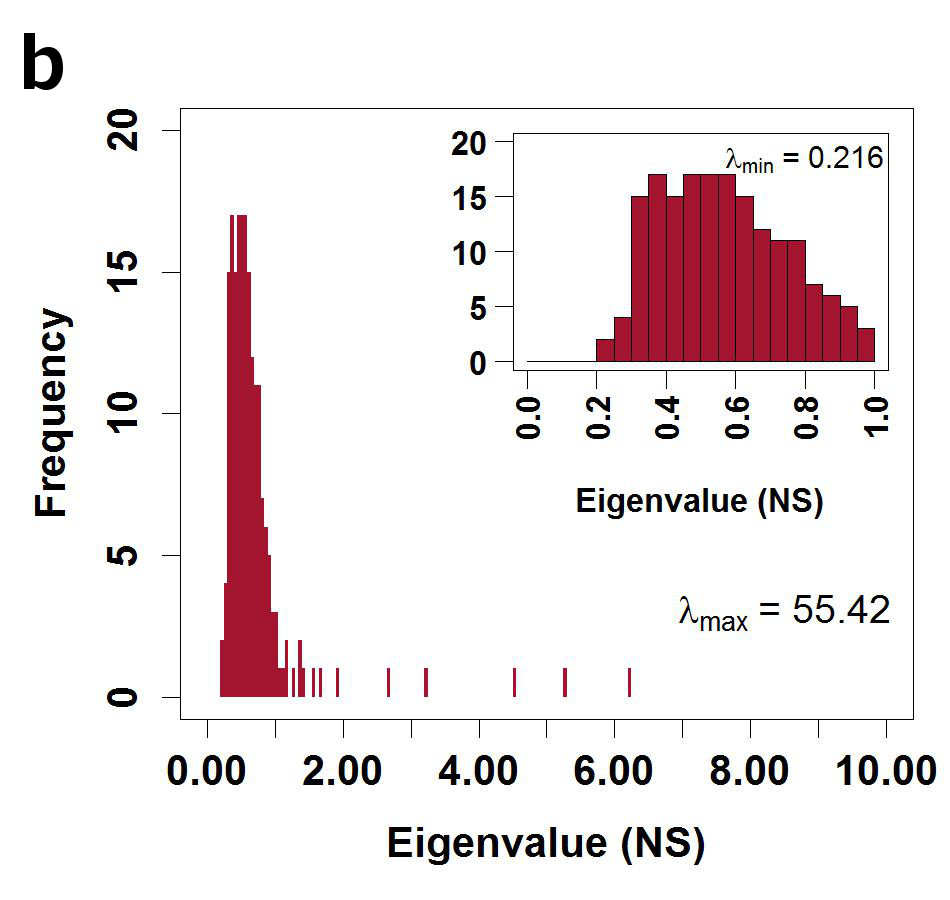}
\includegraphics[width=0.33\linewidth]{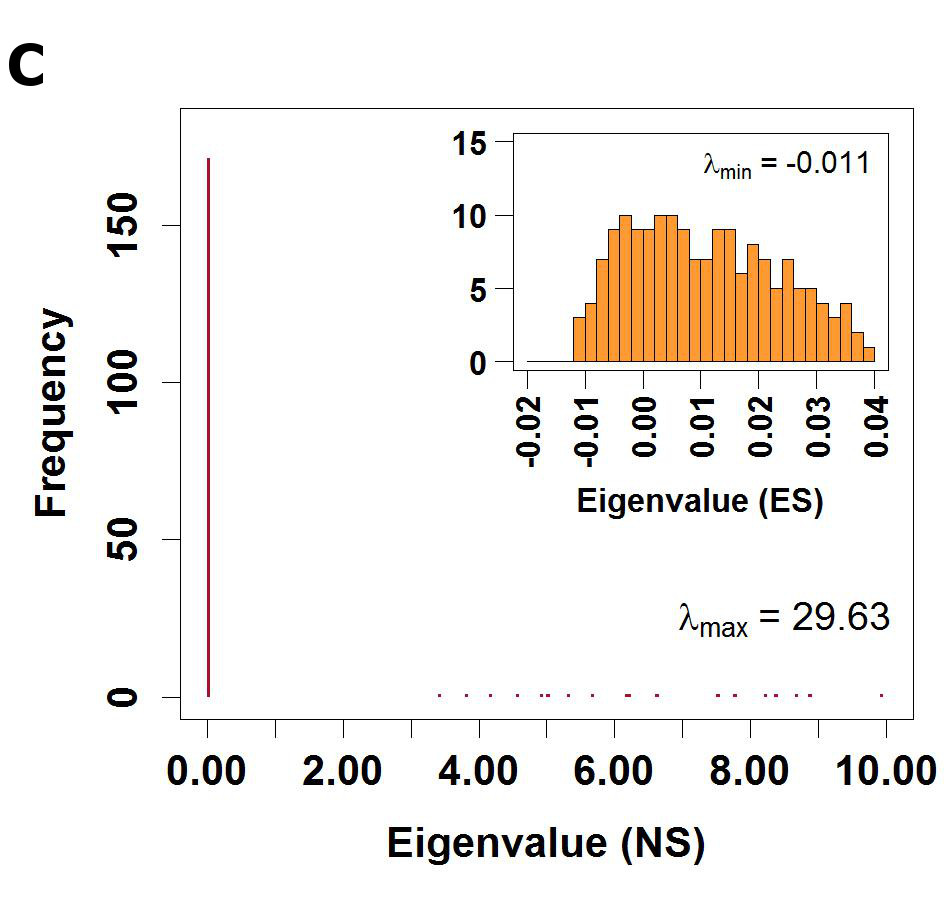}
\includegraphics[width=0.33\linewidth]{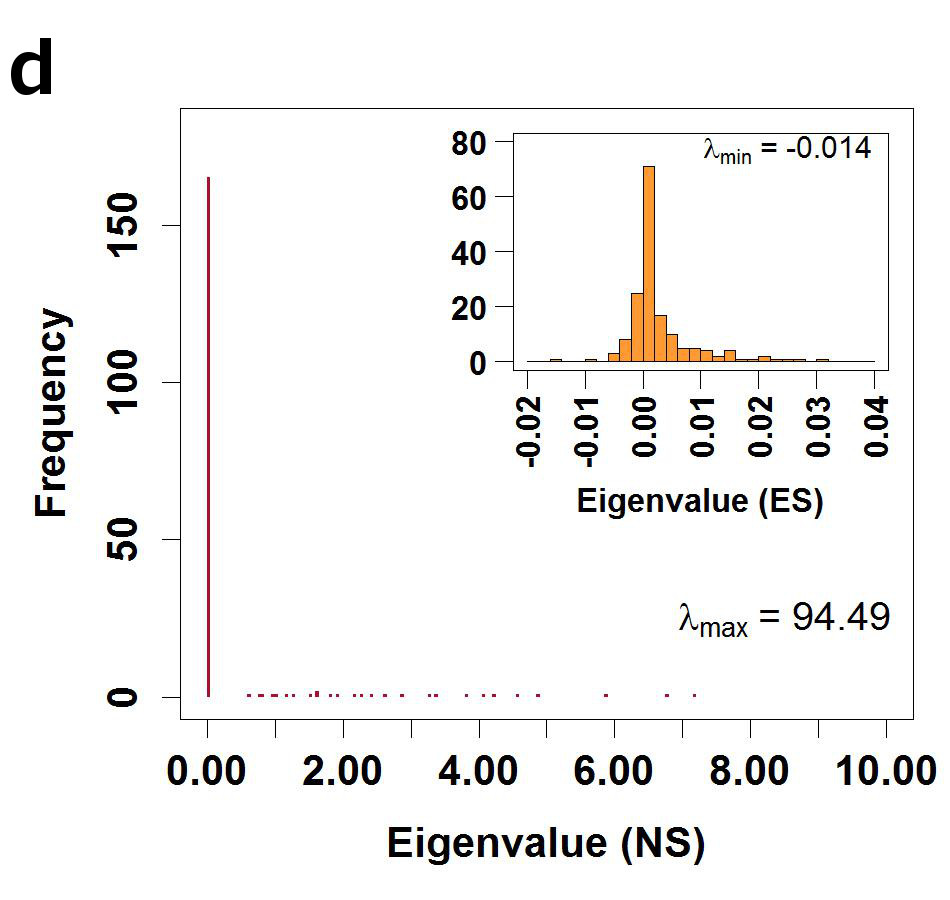}
\caption{\textbf{USA market correlations and eigen spectra.} (\textbf{a}) Schematic diagram showing four sample correlation matrices for different time-epochs of $M=200$ days ending on 30-10-2000, 10-06-2002, 03-08-2009 and 16-12-2011; the correlation structure varies with time -- the market has varying degrees of mean correlation over time. At times one can see that there are strong correlations within certain sectors and anti-correlations with respect to the others; other times, all the sectors are correlated, with the mean market correlation very high (critical periods). The sectoral abbreviations are as follows: 
\textbf{CD}--Consumer Discretionary; 
\textbf{CS}--Consumer Staples;       
\textbf{EG}--Energy;                 
\textbf{FN}--Financials;             
\textbf{HC}--Health Care;            
\textbf{ID}--Industrials;           
\textbf{IT}--Information Technology; 
\textbf{MT}--Materials;              
\textbf{TC}--Telecommunication Services; and             
\textbf{UT}--Utilities.  
(\textbf{b}) Eigenvalue spectrum of the correlation matrix, evaluated for the \textit{long} time series of returns for the entire period of $T=8067$ days, with the maximum eigenvalue of the normal spectrum $\lambda_{max}= 55.42$. Inset shows the empirical Mar\u{c}enko-Pastur distribution \cite{Marcenko1967}, with the smallest eigenvalue of the normal spectrum $\lambda_{min}= 0.216$. (\textbf{c}) Non-critical (normal) period eigenspectrum of the correlation matrix, evaluated for the \textit{short} time series of returns for the time-window of $M=20$ days ending on 08-07-1985, with the maximum eigenvalue of the normal spectrum $\lambda_{max}= 29.63$. Inset: Emerging spectrum using power map technique ($\epsilon= 0.01$) is deformed semi-circular, with the smallest eigenvalue of the emerging spectrum $\lambda_{min}= -0.011$. (\textbf{d}) Critical (crash) period eigenspectrum of the correlation matrix, evaluated for the \textit{short} time series of returns for the time-window of $M=20$ days ending on 15-09-2008, with the maximum eigenvalue of the normal spectrum  $\lambda_{max}= 94.49$. Inset: Emerging spectrum using power map technique ($\epsilon= 0.01$) is Lorentzian, with the smallest eigenvalue of the emerging spectrum $\lambda_{min}= -0.014$. }
\label{fig:correlation_eigenspectra}
\end{figure}

In Figure~\ref{fig:correlation_eigenspectra} (\textbf{a}), four correlation structures with $M=200$ (non-singular matrices) are shown for the USA market with $N=194$ stocks, for the periods ending on 30-10-2000, 10-06-2002, 03-08-2009 and 16-12-2011; evidently, the correlation structure varies with time -- the market has highly correlated structure during the turbulent period (16-12-2011), and an interesting  structure mixed with correlations and anti-correlations during a relatively calm period (30-10-2000), with the mean market correlation $\mu (t)$ varying over time. 
In Figure~\ref{fig:correlation_eigenspectra} (\textbf{b}), the eigenvalue spectrum of the correlation matrix, evaluated for the \textit{long} time series of returns for the entire period of $T=8067$ days, is shown with the maximum eigenvalue of the normal spectrum $\lambda_{max}= 55.42$ (not shown). The Inset of Figure~\ref{fig:correlation_eigenspectra} (\textbf{b}) shows the empirical Mar\u{c}enko-Pastur distribution \cite{Marcenko1967}, with the smallest eigenvalue of the normal spectrum $\lambda_{min}= 0.22$. The maximum eigenvalue $\lambda_{max}$ essentially captures the mean correlation ($\mu$) in the market. Instead of working with a long time series to determine the correlation matrix for $N=194$ stocks, if we work with a \textit{short} time series epoch of  $M=20$ days, then the correlation matrix would be singular and the eigenvalue spectra would have $(N-M+1)=175$ zero eigenvalues (see Figure \ref{fig:correlation_eigenspectra} (\textbf{c-d})).

For the short time-epochs of $M=20$ days, following the methodology of the power map by Vinayak et al. \cite{Vinayak2013}, we now give a non-linear distortion to each cross-correlation matrix within an epoch:
$ C_{ij} \rightarrow (\mathrm{sign} ~~C_{ij}) |C_{ij}|^{1+\epsilon}$, where $\epsilon= 0.01$.
This gives rise to an ``emerging spectrum'' of eigenvalues. The Insets of Figure \ref{fig:correlation_eigenspectra} (\textbf{c}) and (\textbf{d}) show the emerging spectra, which are considerably different from the non-distorted or normal spectrum.  Notably, some of the eigenvalues are now negative! 
We further observe that the statistical properties of the emerging spectra get affected by market turbulence -- the emerging spectrum is distorted semi-circular when the market is normal, while it is Lorentzian when the market is critical (turbulent) with very strong correlations.

\begin{figure}
\centering
	\includegraphics[width=0.48\linewidth]{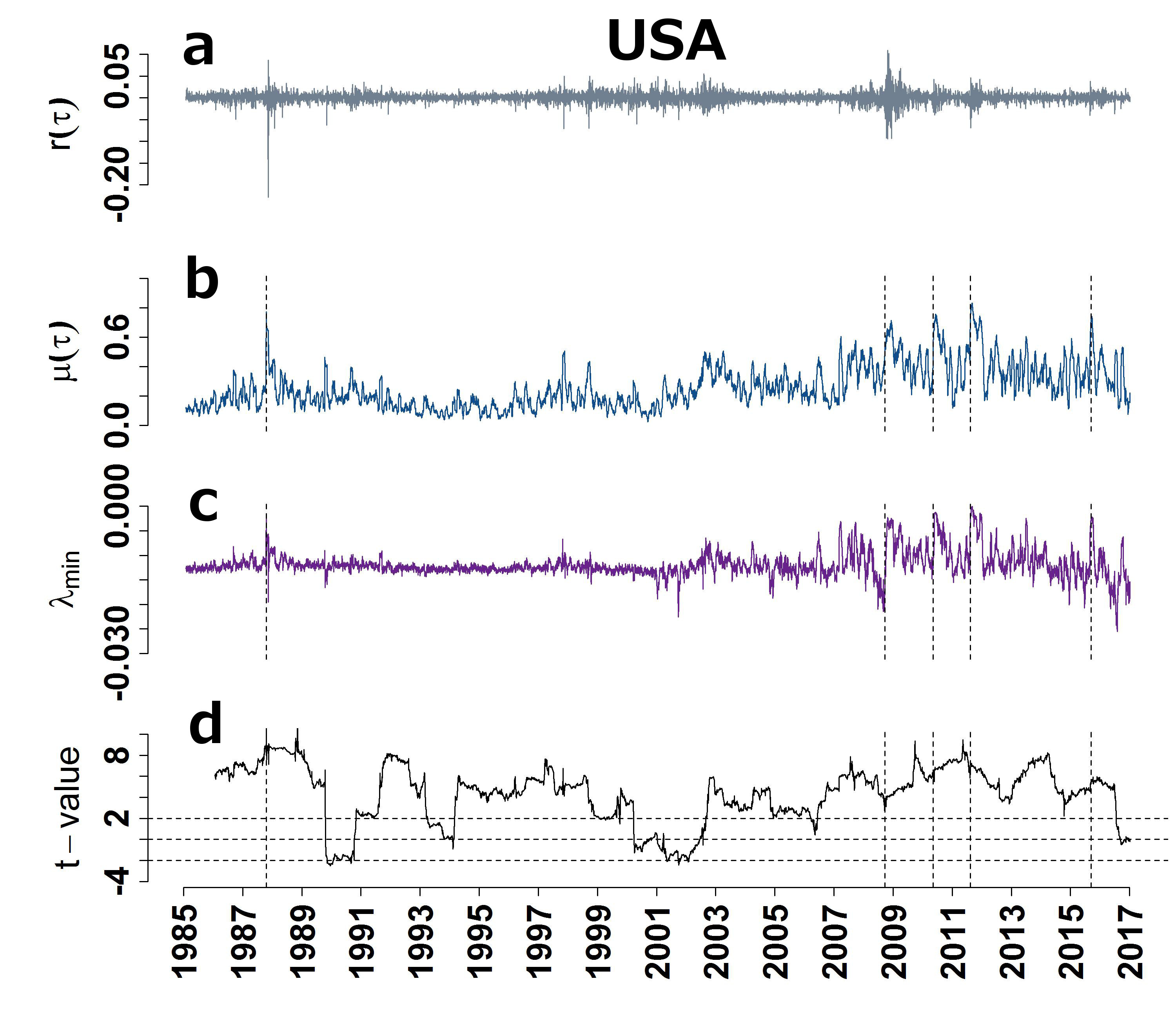}
	\includegraphics[width=0.48\linewidth]{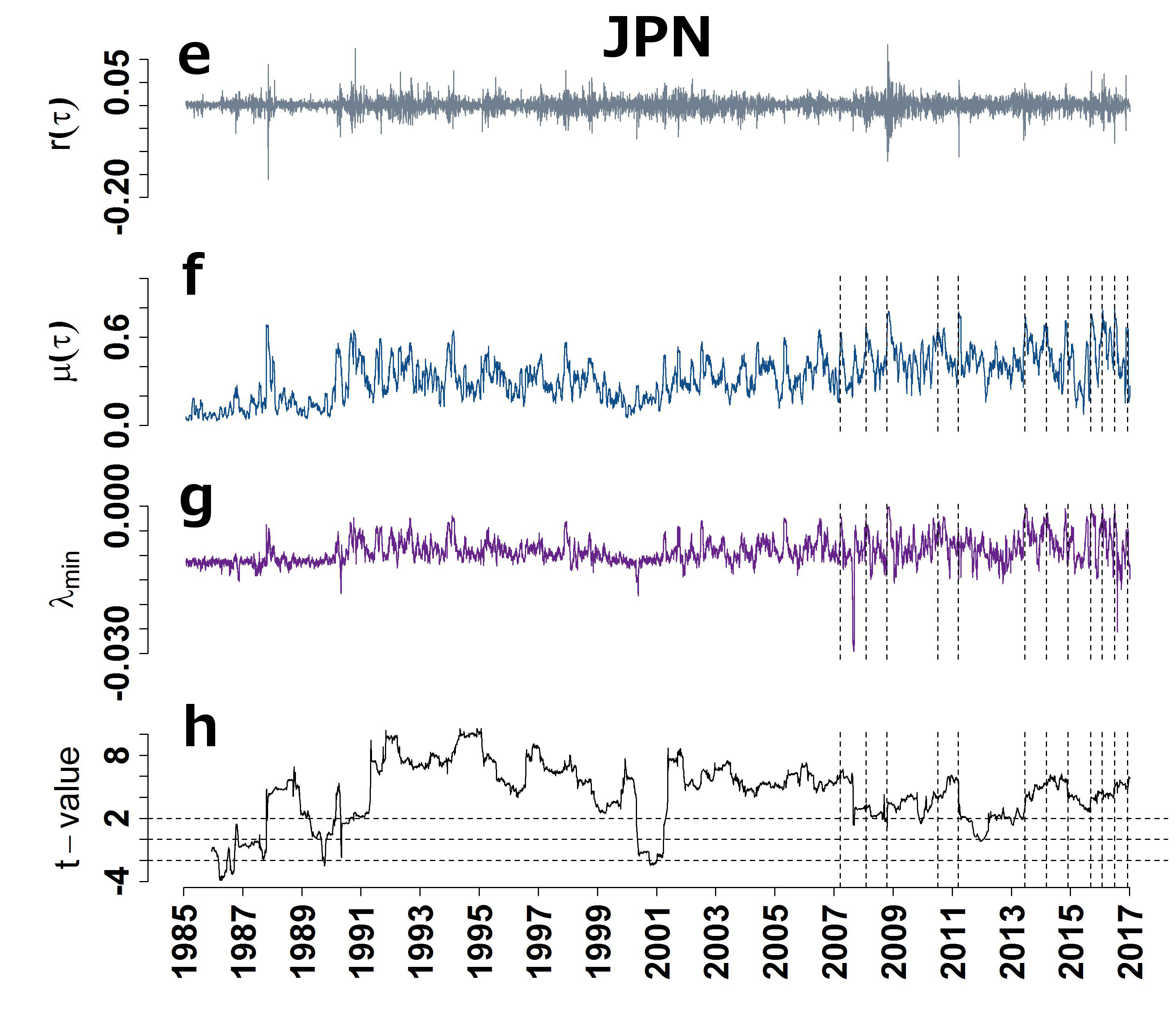}
\caption{\textbf{Comparison of the index returns, mean market correlations and eigenspectral properties.} (\textbf{a-d}) USA and and (\textbf{e-h}) JPN: (i) market returns $r(t)$, (ii) mean market correlation $\mu (t)$, (iii) smallest eigenvalue of the emerging spectrum ($\lambda_{min}$), and (iv) t-value of the t-test, which tests the statistical effect over the lag-1 smallest eigenvalue $\lambda_{min}(t-1)$  on the mean market correlation $\mu(t)$ (described in Methods section). The mean of the correlation coefficients and the smallest eigenvalue in the emerging spectra are correlated to a large extent. Notably, the smallest eigenvalue behaves differently (sharply rising \textit{or} falling) at the same time when the mean market correlation  is very high (crash). The vertical dashed lines correspond to the major crashes, which brewed due to internal market reactions (as confirmed by outlier test in Figure~\ref{fig:spectral_properties} and listed in Table~\ref{table:crashes}). Note that for the USA, from 2001 onward, the smallest eigenvalue seems to indicate that the financial market has become more turbulent. Similarly for Japan, the nature of the market has changed from 1990 onward.}
\label{fig:crises}
\end{figure}

Figure \ref{fig:crises} shows for the USA (\textbf{a-d}) and JPN (\textbf{e-h}): (i) market returns $r(t)$, (ii) mean market correlation $\mu (t)$, (iii) smallest eigenvalue of the emerging spectrum ($\lambda_{min}$), and (iv) t-value of the t-test, which tests if lag-1 smallest eigenvalue $\lambda_{min}(t-1)$ has statistical effect over mean market correlation $\mu(t)$ (described in Methods section). Using the GARCH($p,q$) models for volatility modeling (given in Methods section) on all the three variables (especially the smallest eigenvalue of the emerging spectrum), it is evident that for the USA, from 2001 onward, the financial market has become more turbulent. Similarly for Japan, the nature of the market has changed from 1990 onward. The lag-1 smallest eigenvalue $\lambda_{min}(t-1)$ seems to pick up (statistically significant at $2\sigma$ levels or higher) for most of the time, in ahead by one-day, the signal of how correlated the market would be. This feature could shed light and be exploited in designing market strategies, etc. The only periods when the $\lambda_{min}(t-1)$ fails to detect, are the broad periods (1990-91, 2000-02, etc.) which act like as bubbles or anomalies. However, the ``Dot-com" bubble can be treated as an anomaly, since only a single industry was at crisis. Thus, smallest eigenvalue of the emerging spectrum can be effectively used for the characterization of market crashes and as a signal for market turbulence.

\begin{figure}
\centering
	\includegraphics[width=0.95\linewidth]{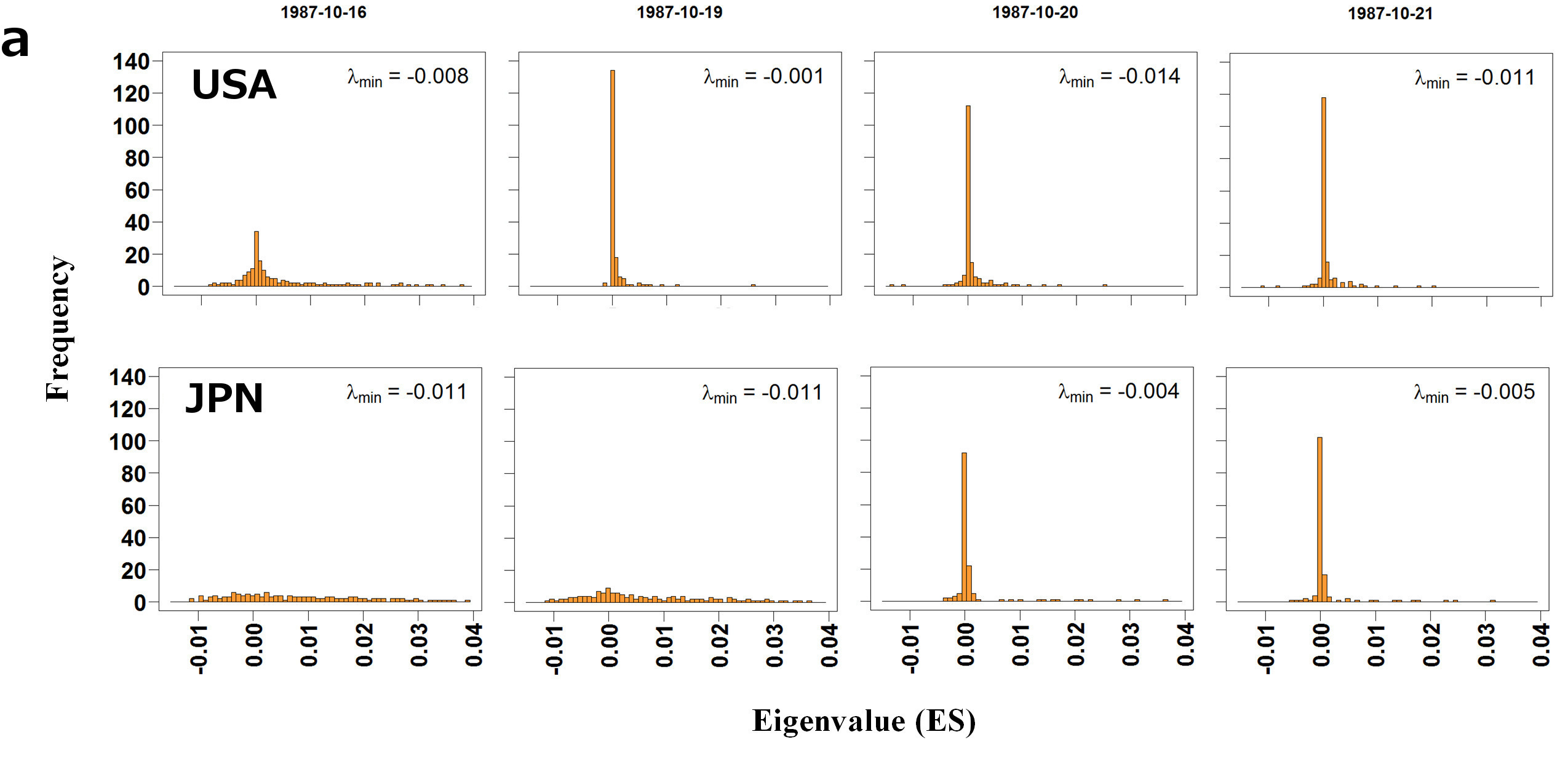}\\
	\includegraphics[width=0.48\linewidth]{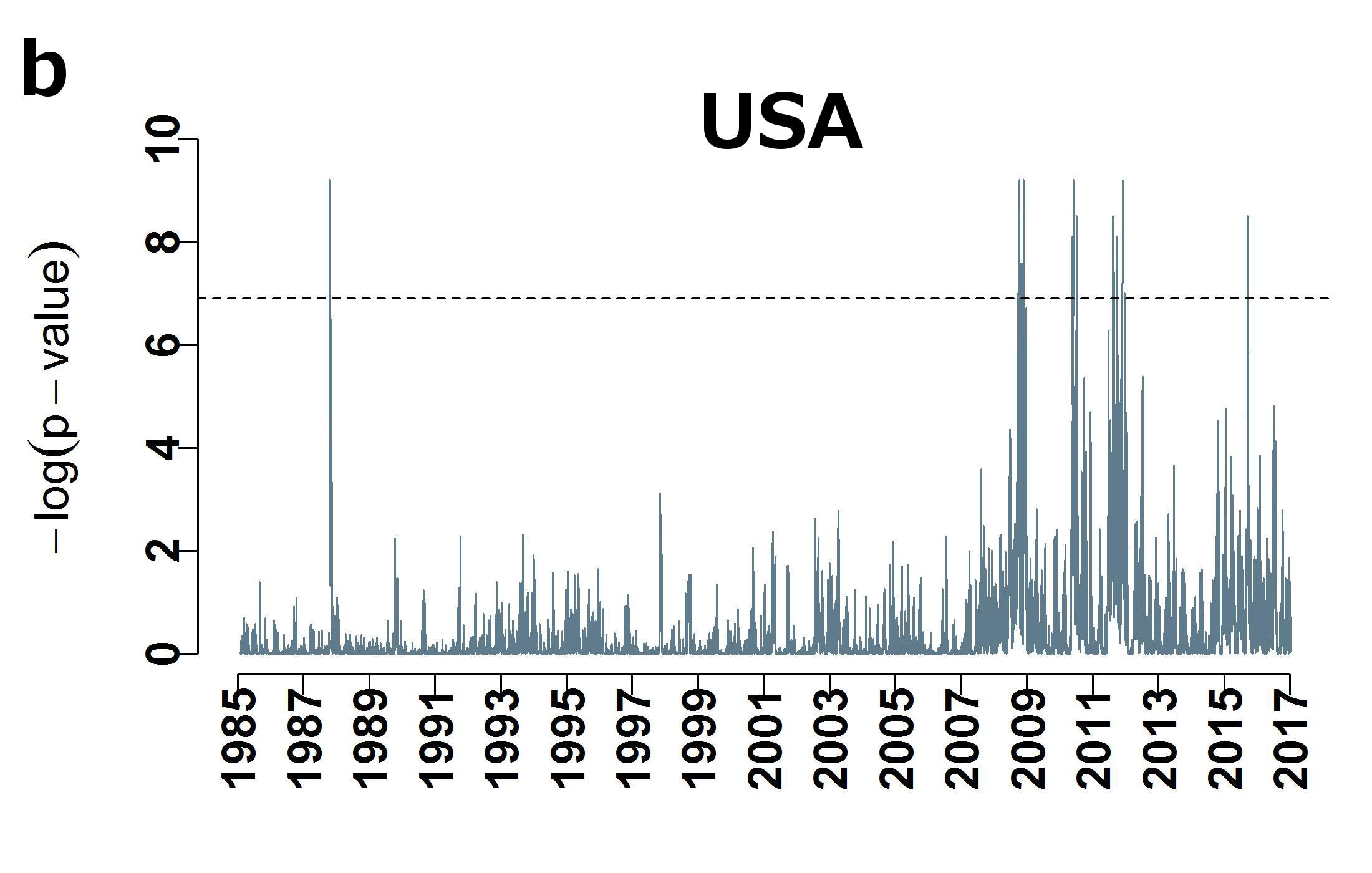}
	\includegraphics[width=0.48\linewidth]{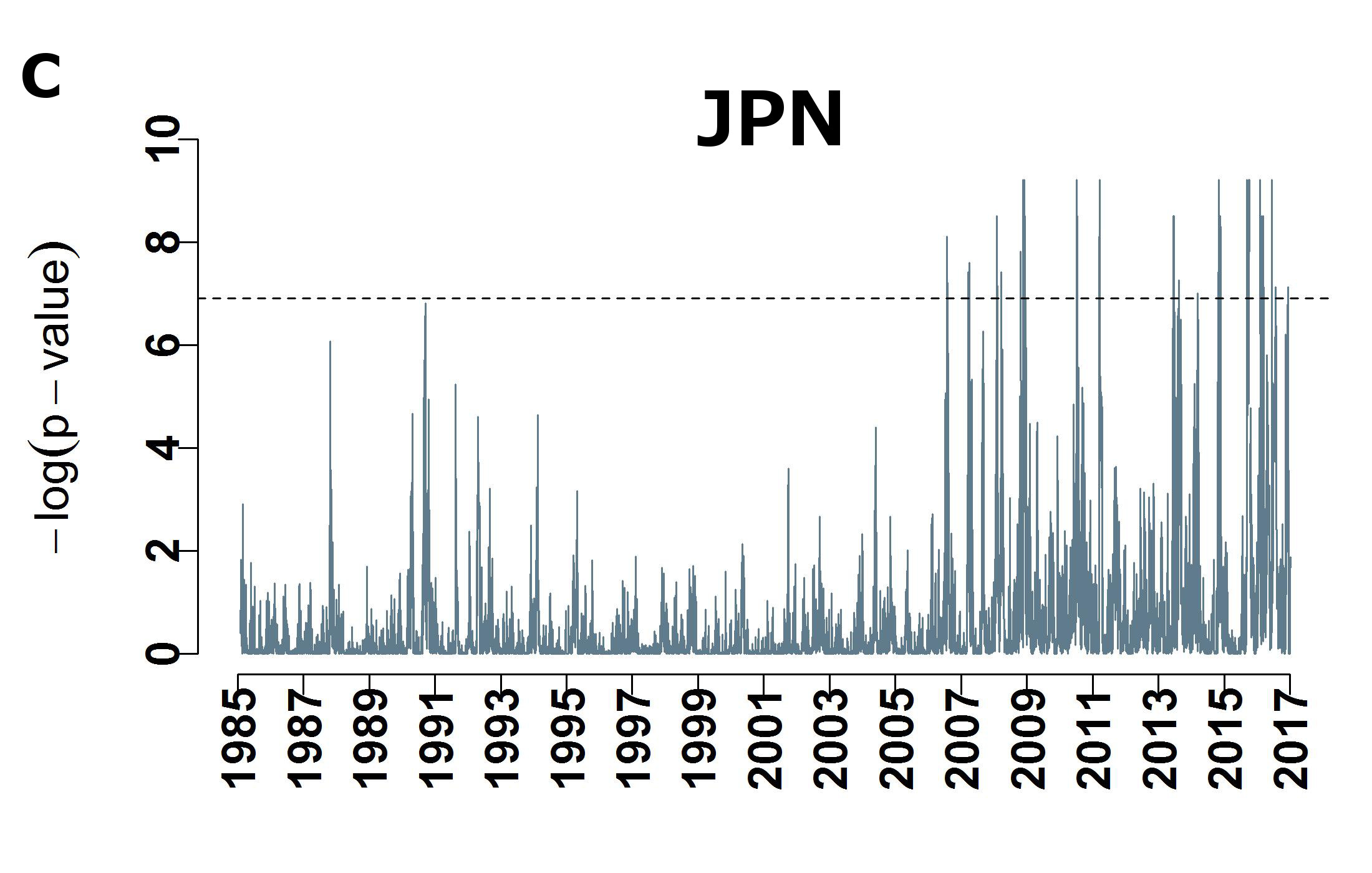}\\
\caption{\textbf{Eigenspectra near catastrophic events and characterisation.} (\textbf{a}) Comparison of the shape of the eigenspectra for USA and Japan, around the Black Monday crash of 19-10-1987. (\textbf{b-c}) The results of the outlier test as a plot of the -log(p-value), so that if the p-value is less than 0.001 (significant at 0.1\% level), then only we reject the null hypothesis that the smallest eigenvalue of the emerging spectrum ($\lambda_{min}$) is not an outlier. The results are shown for USA (\textbf{b}) and JPN (\textbf{c}). It is noteworthy that at 0.1\% level, $\lambda_{min}$ becomes an outlier in USA for the Black Monday crash, sub-prime global crash, european sovereign debt crisis and Chinese stock-market crisis. During Black Monday crash, the p-value does not become significant 0.1\% level. On the other hand, during the Fukushima disaster, the $\lambda_{min}$ becomes an outlier for JPN but not for the USA at the same date. There is notably a lead-lag effect of the crashes across the globe, as can be examined through the behavior of $\lambda_{min}$.}
\label{fig:spectral_properties}
\end{figure}

Figure~\ref{fig:spectral_properties} shows that the shapes of the emerging spectra reflect the turbulence in the market. When the market changes from normal to critical periods, the shape (distribution) of the emerging spectra changes from distorted semi-circular (or similar to Mar\u{c}enko-Pastur) to Lorentzian. Also, the smallest eigenvalue of the emerging spectrum ($\lambda_{min}$) seems to be an outlier whenever the crash brews due to some internal effects. We test whether the smallest eigenvalue of the emerging spectrum ($\lambda_{min}$) is an outlier or not, using the Silverman method \cite{Silverman} (outlined in the Methods section). We find that most of the recent major USA or JPN crashes are due to internal market reactions. The list of the major crashes is given in Table~\ref{table:crashes}, and the corresponding crashes are also indicated in Figure~\ref{fig:crises}. The Black Monday crash is one of the biggest crashes that occurred, and from the behavior of the $\lambda_{min}$, it appears to be due to an internal reaction in USA and an external shock in JPN. There is obviously a delay between the two markets, and the JPN markets reacts after a few days. Thus, the $\lambda_{min}$ could be used as an indicator to study the lead-lag effects in the markets.

\begin{table}[]
\centering
\caption{List of crashes and their characterization}
\begin{tabular}{|l|l|l|l|}
\hline
\textbf{Period}  & \textbf{Region} & \textbf{Affected} & \textbf{Crisis}                                                                 \\ \hline
\textbf{1987}    & Global          & USA               & Black Monday                                                                    \\ \hline
\textbf{2007-09} & Global          & USA, JPN          & \begin{tabular}[c]{@{}l@{}}Subprime Global Financial Crisis\end{tabular} \\ \hline
\textbf{2010-11} & Europe          & USA, JPN          & \begin{tabular}[c]{@{}l@{}}European sovereign debt crisis\end{tabular}       \\ \hline
\textbf{2011}    & Asia            & JPN               & Fukushima Disaster                                                              \\ \hline
\textbf{2013-14} & Europe          & JPN               & Russian Ruble Crisis                                                            \\ \hline
\textbf{2015}  & Asia          & JPN, USA               & Chinese Stock Market Crisis                                                            \\ \hline
\textbf{2016}    & Asia            & JPN          & \begin{tabular}[c]{@{}l@{}}Bank of Japan Policy Crisis\end{tabular}          \\ \hline
\end{tabular}
\label{table:crashes}
\end{table}

\section*{Summary and discussions}

In this paper, we studied the statistical properties of the emerging spectra and showed for the first time that: (i) the shape of the emerging spectrum reflects the market turbulence, and (ii) the smallest eigenvalue of the emerging spectrum may be able to statistically distinguish the principal nature of a market turbulence or crisis -- internal reaction or external shock. Further, when we ran a linear regression model for the mean market cross-correlation as function of the time-lagged smallest eigenvalue, we found that the two variables have statistically significant correlation. 
The surprising and far reaching result we find is that the smallest eigenvalue (of the emerging spectrum) does not only anti-correlate trivially with the largest one, and thus with the average correlation, but in certain instabilities  the anti-correlation turns into positive correlation. This change can reasonably be associated to the question whether a crash is associated to intrinsic market conditions (e.g., a bubble) or to external events (e.g., the Fukushima meltdown). Furthermore, the time-lagged correlation between the smallest eigenvalue and the highest one is not always very small and thus an indicator function may exist. 
These resuls are of deep significance for the undertstanding of financial markets but beyond that open a new window to the exploration of other complex systems that display catastrophic instabilities.

\section*{Methods}
\subsection*{Data Description}
\label{Sec:Materials}
We have used the adjusted closure price time series from the Yahoo finance database \cite{Yahoo_finance}, for two countries: United States of America (USA) S\&P-500 index and Japan (JPN) Nikkei-225 index, for the period 02-01-1985 to 30-12-2016, and for the corresponding stocks as follows:
\begin{itemize}
\item USA --- 02-01-1985 to 30-12-2016 ($T=8068$ days); Number of stocks $N=194$;
\item JPN --- 04-01-1985 to 30-12-2016 ($T=7998$ days); Number of stocks $N=165$,
\end{itemize}
\noindent where we have included the stocks which are present in the index for the entire duration. The list of stocks (along with the sectors) for the two markets are given in the Tables \ref{USA_Table} and \ref{JPN_Table} in Supplementary information.


\subsection*{Cross-correlation Matrix} 
Returns series are constructed as
$r_i( t ) = \ln P_i( t )- \ln P_i( t -1)$, where $P_i(t )$ is the adjusted closure price of stock $i$ in day $t$. Then the equal time Pearson correlation coefficients between stocks $i$ and $j$ is defined as
 $ C_{ij} (t) = (\langle r_i r_j \rangle - \langle r_i \rangle \langle r_j \rangle)/\sigma_i\sigma_j$, 
where $\langle...\rangle$ represents the expectation computed over the time-epochs of size $M$ and the day ending on $t$, and $\sigma_k$ represents standard deviation of the $k$-th stock evaluated for the same time-epochs. We  use $\boldsymbol{C} (t)$ to denote the return correlation matrix for the time-epochs ending on day $t$. 


\subsection*{Test for Outliers}

Silverman \cite{Silverman} proposed a technique for using kernel density
estimates to investigate the number of modes in a population. This
technique can be used to identify if there is an outlier in the
data. If there exists one or more outliers, then we can say that outliers are
generating a separate (and distinct) minor mode other than the major mode, and
the distribution would have more than a single mode. On the other hand, if there
does not exist any outlier, then the distribution would have only one
mode and all samples would be generated from a uni-modal
distribution. 

We wish to find if the smallest eigenvalue ($\lambda_{min}$) is
an outlier or not. So we set up the null hypothesis as -- $\lambda_{min}$ is not an outlier, and the alternative hypothesis as --
 $\lambda_{min}$ is an outlier.  Further, as indicated in the Lorentzian distribution of Figure \ref{fig:spectral_properties} (\textbf{a}), we are
interested in the lower part of the emerging eigenspectra $\lambda$ which has the $\lambda_{min}$. So, we
consider only the conditional distribution of $\lambda$ given the
median eigenvalue $\lambda_{(m)}$, i.e., $f(\lambda | \lambda
<\lambda_{(m)})$. Hence, the equivalent null hypothesis can be stated as
$f(\lambda | \lambda <\lambda_{(m)})$ has a single mode. Mathematically,
$$
H_0: f(\lambda | \lambda <\lambda_{(m)}) \textrm{ has exactly one mode}.
$$
The alternative hypothesis is 
$$
H_A: f(\lambda | \lambda <\lambda_{(m)}) \textrm{ has at least two modes}.
$$
The Silverman test chooses the amount of smoothening automatically. Throughout, we set the level of significance at $0.1\%$
level. That is, if the p-value is less than 0.001, then only we reject the null hypothesis.


\subsection*{Linear regression Model for market correlation on lagged Smallest Eigenvalue}
We consider a linear regression model for $\mu$ (mean market correlation):
$$
\mu(t)=\beta_0+\beta_1\lambda_{min}(t-1)+\beta_2 \lambda_{min}(t-2)+...+\beta_p \lambda_{min}(t-p)+\epsilon(t),
$$
where the $\beta$'s are the coefficients to be estimated, $\epsilon \sim N(0,\sigma^2), ~~ t= 0,1,2,...,T$ is the white noise,  $\mu(t)$ is the mean market correlation at time point $t$, $\lambda_{min}(t-1)$ is the lag-$1$
smallest eigenvalue, $\lambda_{min}(t-2)$ is the lag-$2$
smallest eigenvalue, and $\lambda_{min}(t-p)$ is the lag-$p$
smallest eigenvalue. Here, we choose $p=3$ for our model.

We then perform the t-test, which tests if lag-1 smallest eigenvalue $\lambda_{min}(t-1)$ has statistical effect of mean market correlation 
$\mu(t)$; so, the null hypothesis and the alternate hypothesis can be stated mathematically as: 
$$
H_0: \beta_1=0 ~~~vs ~~~H_A:\beta_1 \neq 0.
$$

The
$t$-value for estimated $\hat{\beta_1}$ is calculated as
$$
t=\frac{\hat{\beta_1}-0}{se(\hat{\beta_1})},
$$
where $se$ is the standard error in statistics. If the value of $|t|>2$, we can say that the last day's SEV ($\lambda_{min}(t-1)$)
has statistically significant effect over today's mean correlation
($\mu(t)$). The value of $t$ itself signifies the strength of the signal. The t-value for estimated $\hat{\beta_1}$ over time is presented in Figure~\ref{fig:crises}.

\subsection*{GARCH($p,q$)}

The ARCH($p$) process \cite{Tsay} is defined by the equation
\begin{equation}
\sigma_t^2=\alpha_0+\alpha_1 x_{t-1}^{2}+...+\alpha_p x_{t-p}^{2} \, ,
\label{archp}
\end{equation}
where the $\{\alpha_0,\alpha_1,...\alpha_p\}$ are positive parameters and $x_{t}$ is a 
random variable with zero mean and variance $\sigma_t^2$, characterized by
a conditional probability distribution 
function $f_t(x)$, which may be chosen as Gaussian. The nature of the memory
of the variance $\sigma_t^2$ is determined by the parameter $p$.

The generalized ARCH process GARCH($p,q$) was introduced 
by Bollerslev \cite{bollerslev,Tsay} and is defined by the equation
\begin{equation}
\sigma_t^2=\alpha_0+\alpha_1 x_{t-1}^{2}+...+\alpha_q x_{t-q}^{2}+\beta_{1}\sigma_{t-1}^{2}+...+\beta_{p}\sigma_{t-p}^{2} \, ,
\label{garchpq}
\end{equation}
where $\{\beta_{1},...,\beta_{p}\}$ are additional control parameters.

The simplest GARCH process is the GARCH(1,1) process, with Gaussian 
conditional probability distribution function, 
\begin{equation}
\sigma_t^2=\alpha_0+\alpha_1 x_{t-1}^{2}+\beta_{1}\sigma_{t-1}^{2} \, .
\label{garch11}
\end{equation}
The random variable $x_{t}$ can be written in term of $\sigma_t$ defining
$x_{t}\equiv\eta_t\sigma_t$,
where $\eta_t$ is a random Gaussian process with zero mean and unit variance.
One can rewrite Eq. \ref{garch11} as a random multiplicative process
\begin{equation}
\sigma_t^2=\alpha_0+(\alpha_1 \eta_{t-1}^{2}+\beta_{1})\sigma_{t-1}^{2} \, .
\label{garch12}
\end{equation}
\begin{figure}[h]
\centering
\includegraphics[width=0.4\linewidth]{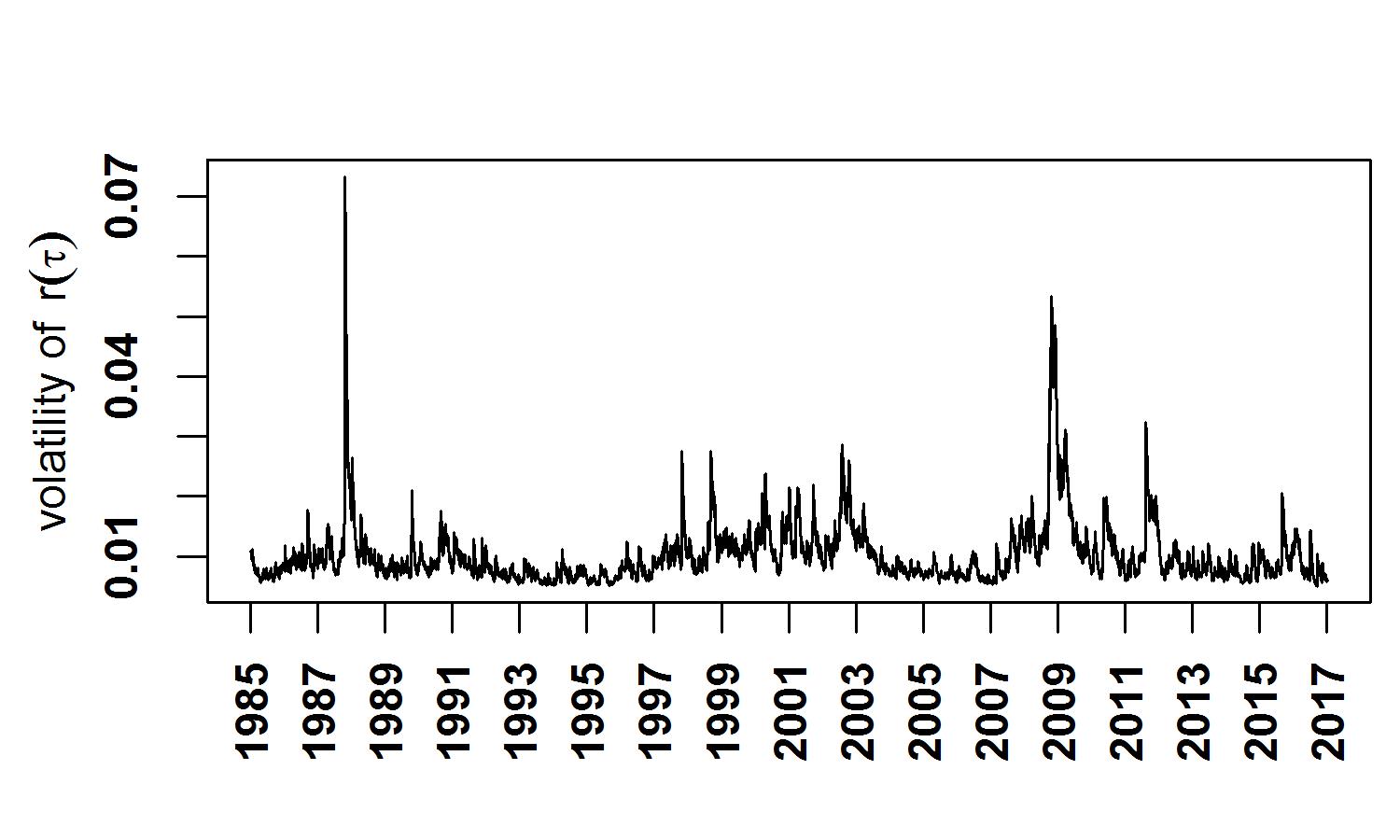}
\includegraphics[width=0.4\linewidth]{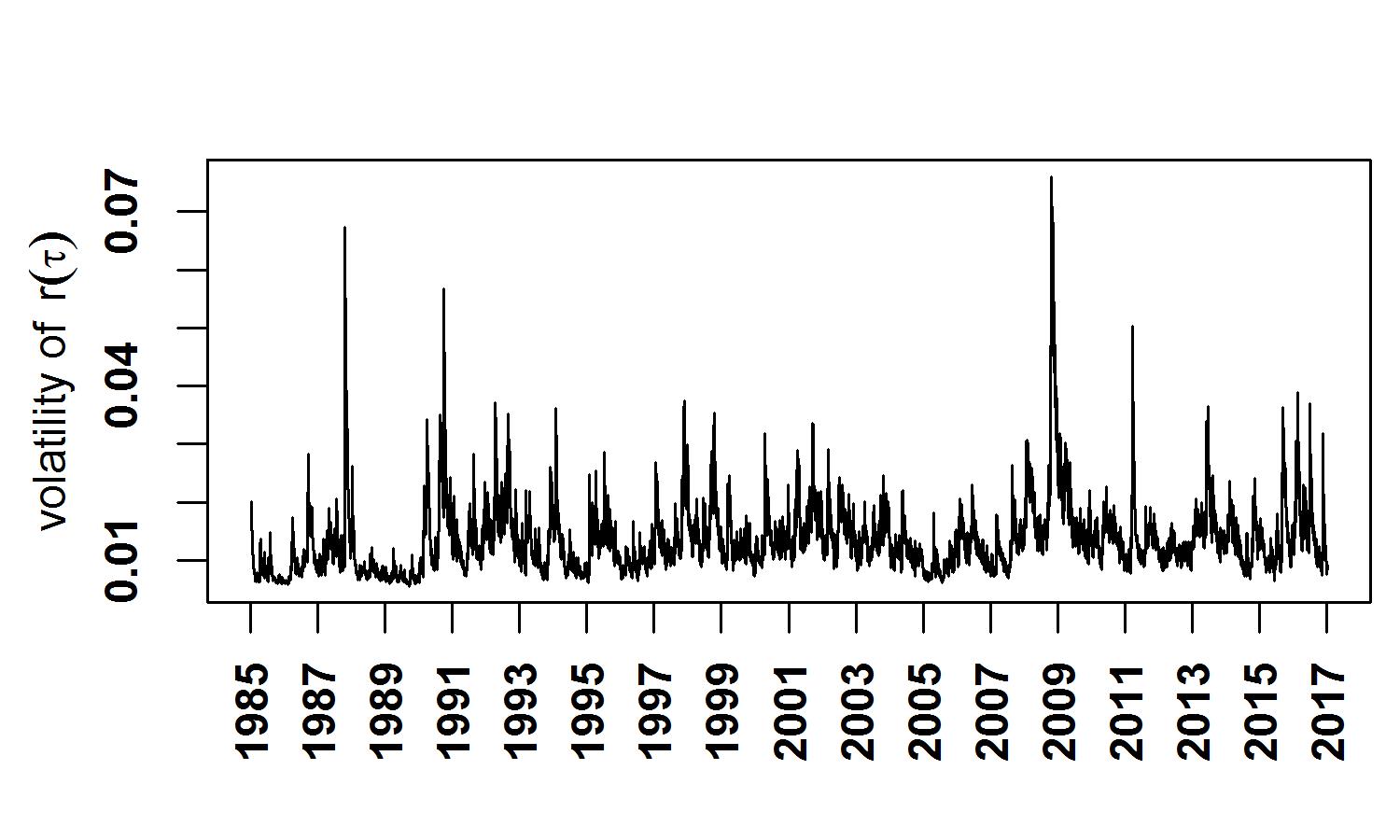}\\
\includegraphics[width=0.4\linewidth]{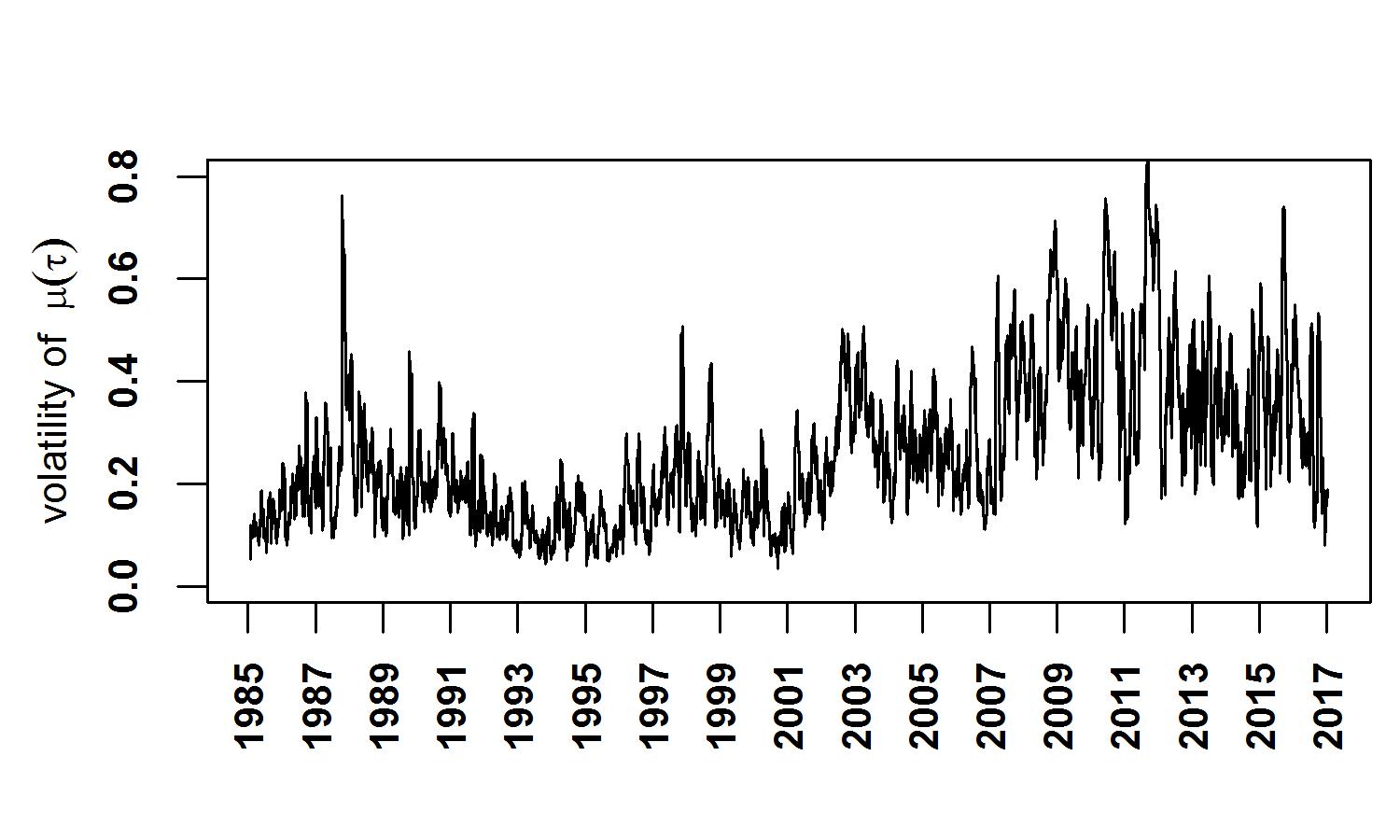}
\includegraphics[width=0.4\linewidth]{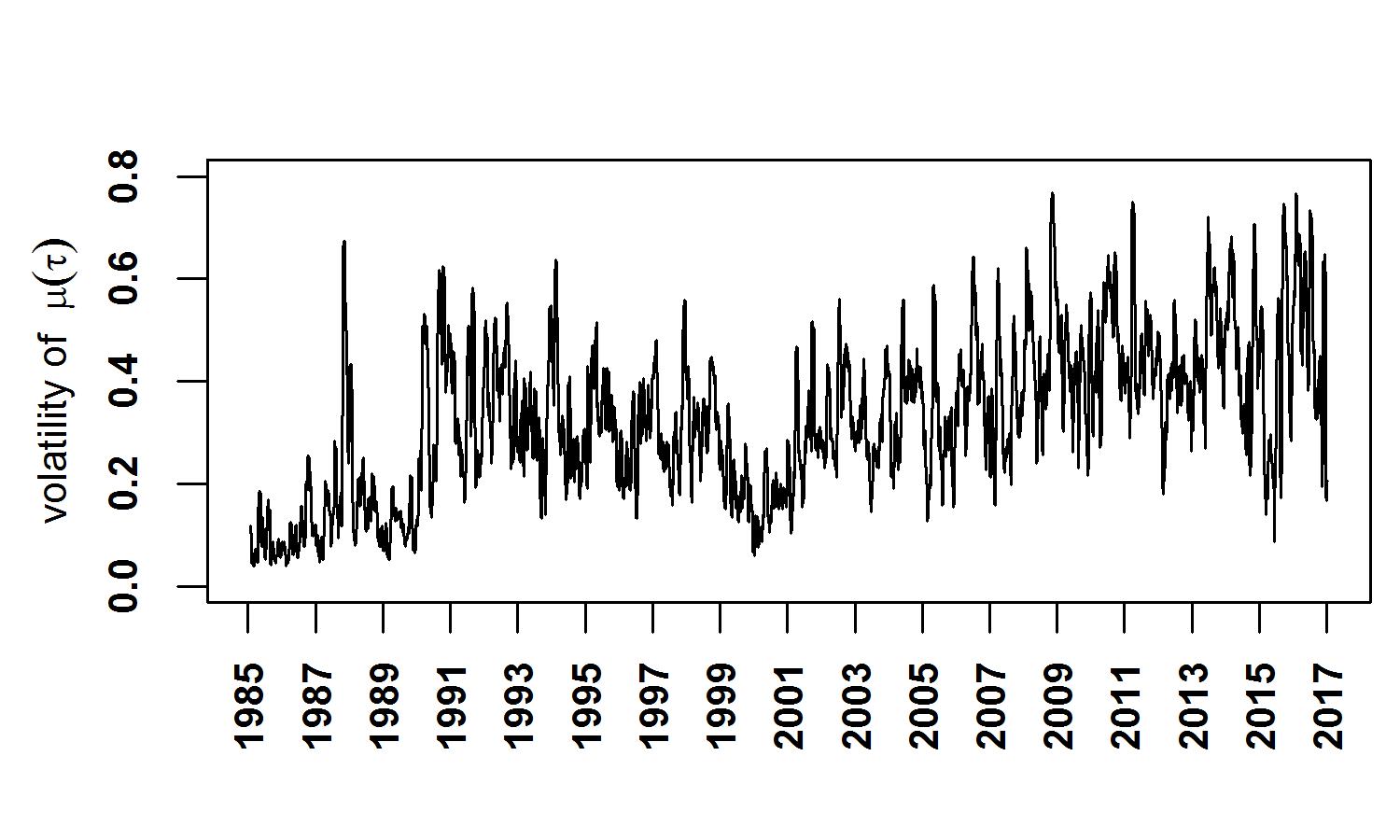}\\
\includegraphics[width=0.4\linewidth]{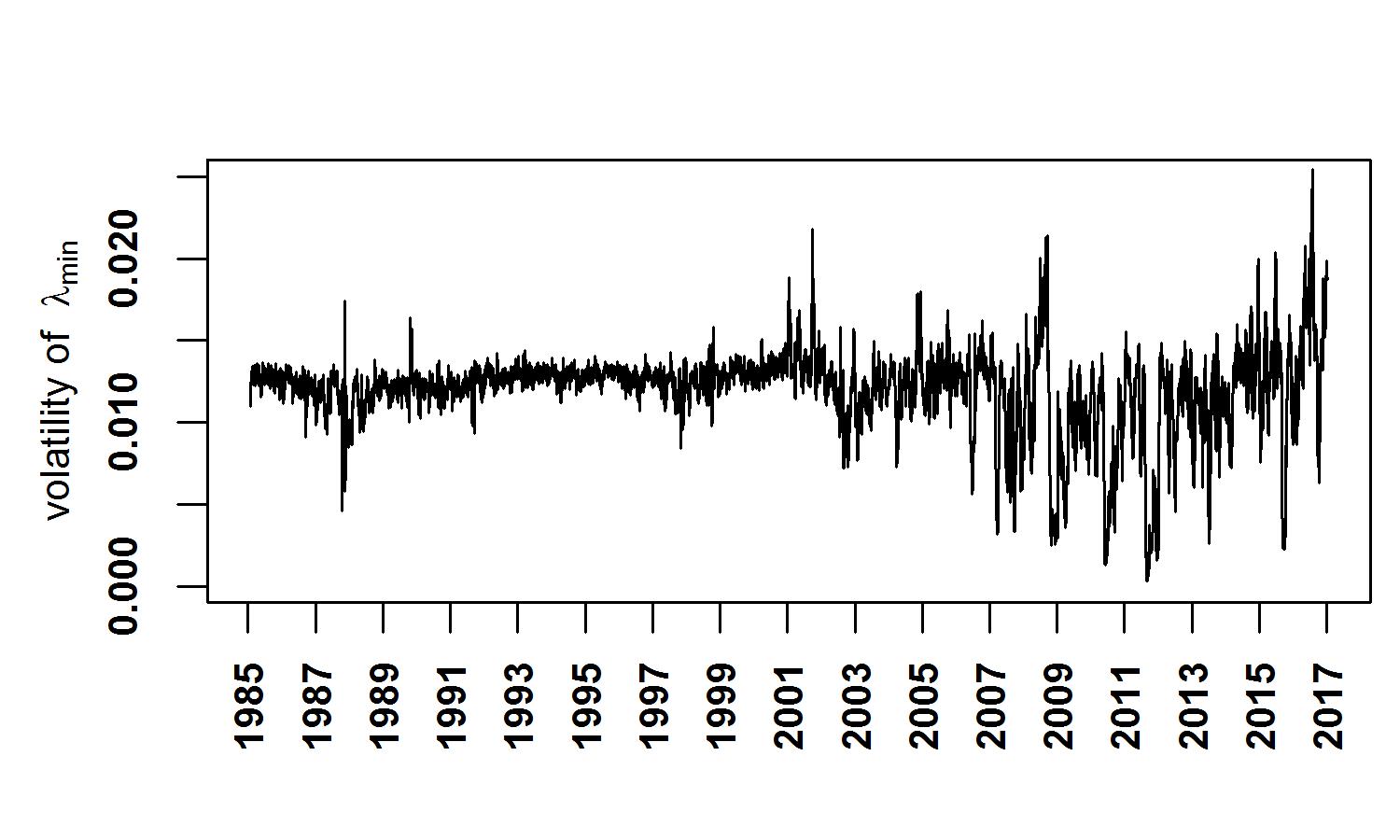}
\includegraphics[width=0.4\linewidth]{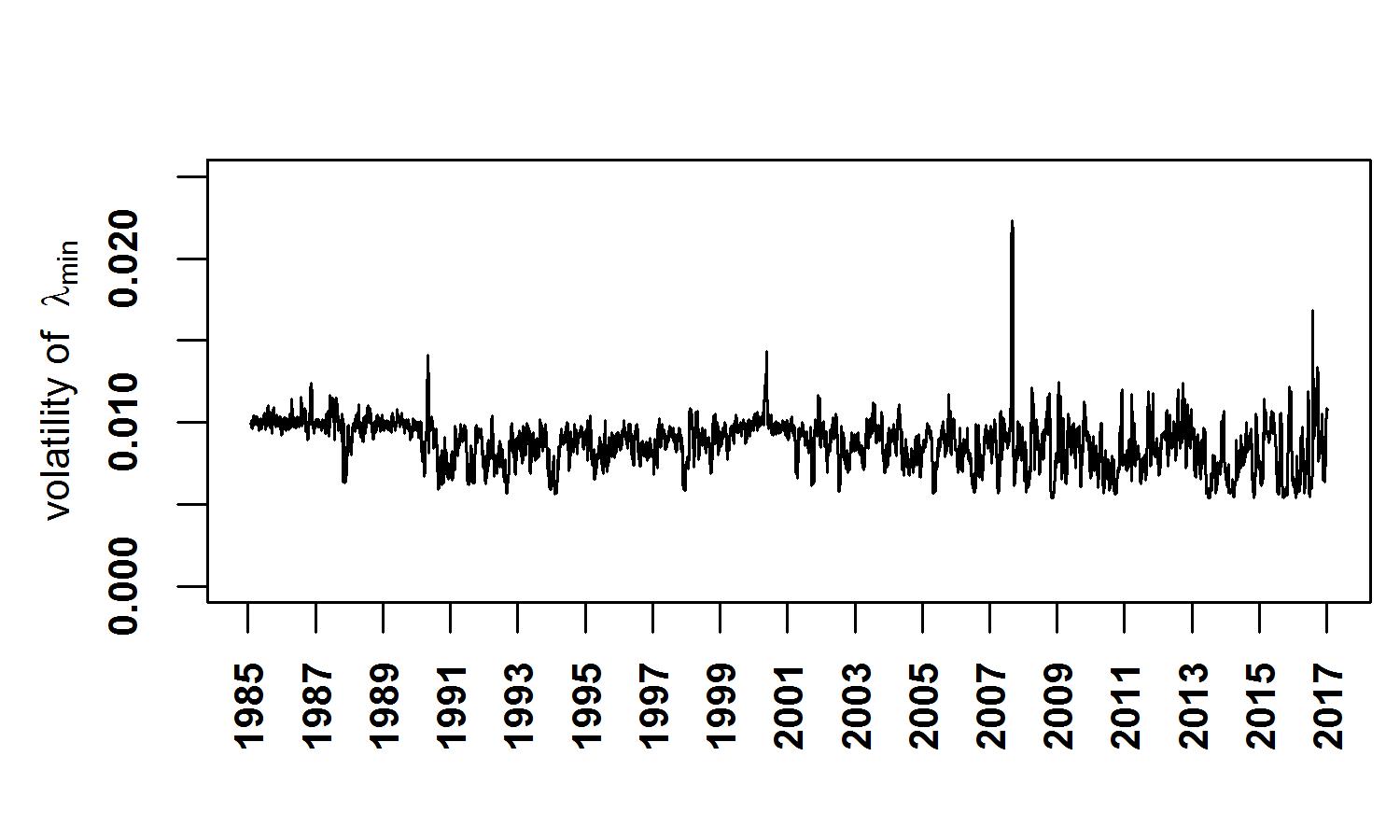}
\caption{\textbf{Volatility estimation using GARCH(1,1) model for USA (Left) and JPN (Right).} (\textbf{Top}) Using market index return, (\textbf{Middle}) using mean of market correlation, and (\textbf{Bottom}) using smallest eigenvalue.}
\label{fig:GARCH}
\end{figure}

The results for the GARCH(1,1) model fitted on the three variables are shown in the Figure \ref{fig:GARCH}.



\section*{Acknowledgements}

The authors are grateful to Anindya S. Chakrabarti,  Sujoy Chakravarty and Francois Leyvraz for their critical inputs and suggestions.
A.C. and K.S. acknowledge the support by grant number BT/BI/03/004/2003(C) of Govt. of India, Ministry of Science and Technology, Department of Biotechnology, Bioinformatics division, University of Potential Excellence-II grant (Project ID-47) of JNU, New Delhi, and the DST-PURSE grant given to JNU by the Department of Science and Technology, Government of India.
K.S. acknowledges the University Grants Commission (Ministry of Human Research Development, Govt. of India) for her senior research fellowship. S.D. is partially supported by the Infosys grant to CMI. H.K.P. and R.C. are grateful for postdoctoral fellowships provided by UNAM-DGAPA. T.H.S. acknowledges the support grant by CONACyT through Project FRONTERAS 201. A.C., K.S. and T.H.S. acknowledge support from the project UNAM-DGAPA-PAPIIT IG 100616. 

\section*{Author contributions statement}

A.C. and T.H.S. designed research; A.C., K.S., H.K.P., and S.D. performed research;  K.S., H.K.P. and R.C. processed and analyzed data; K.S. prepared all the figures, and  A.C. and T.H.S. wrote the manuscript with input from all authors.

\section*{Supplementary information: Lists of stocks along with sectors}

\begin{table}[h]
\centering
\caption{List of all stocks of USA market (S\&P 500) considered for the analysis. The first column has the serial number,
the second column has the abbreviation, the third column has the full name of the stock, and the
fourth column specifies the sector as given in the S\&P 500.}
\label{USA_Table}
\begin{tabular}{|l|l|l|l|l|}
\hline
\textbf{S.No.} & \textbf{Code} & \textbf{Company Name} & \textbf{Sector} & \textbf{Abbrv} \\ \hline
1 & CMCSA & Comcast Corp. & Consumer Discretionary & CD \\ \hline
2 & DIS & The Walt Disney Company & Consumer Discretionary & CD \\ \hline
3 & F & Ford Motor & Consumer Discretionary & CD \\ \hline
4 & GPC & Genuine Parts & Consumer Discretionary & CD \\ \hline
5 & GPS & Gap Inc. & Consumer Discretionary & CD \\ \hline
6 & GT & Goodyear Tire \& Rubber & Consumer Discretionary & CD \\ \hline
7 & HAS & Hasbro Inc. & Consumer Discretionary & CD \\ \hline
8 & HD & Home Depot & Consumer Discretionary & CD \\ \hline
9 & HRB & Block H\&R & Consumer Discretionary & CD \\ \hline
10 & IPG & Interpublic Group & Consumer Discretionary & CD \\ \hline
11 & JCP & J. C. Penney Company, Inc. & Consumer Discretionary & CD \\ \hline
12 & JWN & Nordstrom & Consumer Discretionary & CD \\ \hline
13 & LEG & Leggett \& Platt & Consumer Discretionary & CD \\ \hline
14 & LEN & Lennar Corp. & Consumer Discretionary & CD \\ \hline
15 & LOW & Lowe's Cos. & Consumer Discretionary & CD \\ \hline
16 & MAT & Mattel Inc. & Consumer Discretionary & CD \\ \hline
17 & MCD & McDonald's Corp. & Consumer Discretionary & CD \\ \hline
18 & NKE & Nike & Consumer Discretionary & CD \\ \hline
19 & SHW & Sherwin-Williams & Consumer Discretionary & CD \\ \hline
20 & TGT & Target Corp. & Consumer Discretionary & CD \\ \hline
21 & VFC & V.F. Corp. & Consumer Discretionary & CD \\ \hline
22 & WHR & Whirlpool Corp. & Consumer Discretionary & CD \\ \hline
23 & ADM & Archer-Daniels-Midland Co & Consumer Staples & CS \\ \hline
24 & AVP & Avon Products, Inc. & Consumer Staples & CS \\ \hline
25 & CAG & Conagra Brands & Consumer Staples & CS \\ \hline
26 & CL & Colgate-Palmolive & Consumer Staples & CS \\ \hline
27 & CPB & Campbell Soup & Consumer Staples & CS \\ \hline
28 & CVS & CVS Health & Consumer Staples & CS \\ \hline
29 & GIS & General Mills & Consumer Staples & CS \\ \hline
30 & HRL & Hormel Foods Corp. & Consumer Staples & CS \\ \hline
31 & HSY & The Hershey Company & Consumer Staples & CS \\ \hline
32 & K & Kellogg Co. & Consumer Staples & CS \\ \hline
33 & KMB & Kimberly-Clark & Consumer Staples & CS \\ \hline
34 & KO & Coca-Cola Company (The) & Consumer Staples & CS \\ \hline
35 & KR & Kroger Co. & Consumer Staples & CS \\ \hline
36 & MKC & McCormick \& Co. & Consumer Staples & CS \\ \hline
37 & MO & Altria Group Inc & Consumer Staples & CS \\ \hline
38 & SYY & Sysco Corp. & Consumer Staples & CS \\ \hline
39 & TAP & Molson Coors Brewing Company & Consumer Staples & CS \\ \hline
40 & TSN & Tyson Foods & Consumer Staples & CS \\ \hline
41 & WMT & Wal-Mart Stores & Consumer Staples & CS \\ \hline
42 & APA & Apache Corporation & Energy & EG \\ \hline
43 & COP & ConocoPhillips & Energy & EG \\ \hline
44 & CVX & Chevron Corp. & Energy & EG \\ \hline
45 & ESV & Ensco plc & Energy & EG \\ \hline
46 & HAL & Halliburton Co. & Energy & EG \\ \hline
47 & HES & Hess Corporation & Energy & EG \\ \hline
48 & HP & Helmerich \& Payne & Energy & EG \\ \hline
49 & MRO & Marathon Oil Corp. & Energy & EG \\ \hline
50 & MUR & Murphy Oil Corporation & Energy & EG \\ \hline

\end{tabular}
\end{table}
\begin{table}[]
\centering
\begin{tabular}{|l|l|l|l|l|}
\hline
51 & NBL & Noble Energy Inc & Energy & EG \\ \hline
52 & NBR & Nabors Industries Ltd. & Energy & EG \\ \hline
53 & SLB & Schlumberger Ltd. & Energy & EG \\ \hline
54 & TSO & Tesoro Corp & Energy & EG \\ \hline
55 & VLO & Valero Energy & Energy & EG \\ \hline
56 & WMB & Williams Cos. & Energy & EG \\ \hline
57 & XOM & Exxon Mobil Corp. & Energy & EG \\ \hline
58 & AFL & AFLAC Inc & Financials & FN \\ \hline
59 & AIG & American International Group, Inc. & Financials & FN \\ \hline
60 & AON & Aon plc & Financials & FN \\ \hline
61 & AXP & American Express Co & Financials & FN \\ \hline
62 & BAC & Bank of America Corp & Financials & FN \\ \hline
63 & BBT & BB\&T Corporation & Financials & FN \\ \hline
64 & BEN & Franklin Resources & Financials & FN \\ \hline
65 & BK & The Bank of New York Mellon Corp. & Financials & FN \\ \hline
66 & C & Citigroup Inc. & Financials & FN \\ \hline
67 & CB & Chubb Limited & Financials & FN \\ \hline
68 & CINF & Cincinnati Financial & Financials & FN \\ \hline
69 & CMA & Comerica Inc. & Financials & FN \\ \hline
70 & EFX & Equifax Inc. & Financials & FN \\ \hline
71 & FHN & First Horizon National Corporation & Financials & FN \\ \hline
72 & HBAN & Huntington Bancshares & Financials & FN \\ \hline
73 & HCN & Welltower Inc. & Financials & FN \\ \hline
74 & HST & Host Hotels \& Resorts, Inc. & Financials & FN \\ \hline
75 & JPM & JPMorgan Chase \& Co. & Financials & FN \\ \hline
76 & L & Loews Corp. & Financials & FN \\ \hline
77 & LM & Legg Mason, Inc. & Financials & FN \\ \hline
78 & LNC & Lincoln National & Financials & FN \\ \hline
79 & LUK & Leucadia National Corp. & Financials & FN \\ \hline
80 & MMC & Marsh \& McLennan & Financials & FN \\ \hline
81 & MTB & M\&T Bank Corp. & Financials & FN \\ \hline
82 & PSA & Public Storage & Financials & FN \\ \hline
83 & SLM & SLM Corporation & Financials & FN \\ \hline
84 & TMK & Torchmark Corp. & Financials & FN \\ \hline
85 & TRV & The Travelers Companies Inc. & Financials & FN \\ \hline
86 & USB & U.S. Bancorp & Financials & FN \\ \hline
87 & VNO & Vornado Realty Trust & Financials & FN \\ \hline
88 & WFC & Wells Fargo & Financials & FN \\ \hline
89 & WY & Weyerhaeuser Corp. & Financials & FN \\ \hline
90 & ZION & Zions Bancorp & Financials & FN \\ \hline
91 & ABT & Abbott Laboratories & Health Care & HC \\ \hline
92 & AET & Aetna Inc & Health Care & HC \\ \hline
93 & AMGN & Amgen Inc & Health Care & HC \\ \hline
94 & BAX & Baxter International Inc. & Health Care & HC \\ \hline
95 & BCR & Bard (C.R.) Inc. & Health Care & HC \\ \hline
96 & BDX & Becton Dickinson & Health Care & HC \\ \hline
97 & BMY & Bristol-Myers Squibb & Health Care & HC \\ \hline
98 & CAH & Cardinal Health Inc. & Health Care & HC \\ \hline
99 & CI & CIGNA Corp. & Health Care & HC \\ \hline
100 & HUM & Humana Inc. & Health Care & HC \\ \hline

\end{tabular}
\end{table}

\begin{table}[]
\centering
\begin{tabular}{|l|l|l|l|l|}
\hline
101 & JNJ & Johnson \& Johnson & Health Care & HC \\ \hline
102 & LLY & Lilly (Eli) \& Co. & Health Care & HC \\ \hline
103 & MDT & Medtronic plc & Health Care & HC \\ \hline
104 & MRK & Merck \& Co. & Health Care & HC \\ \hline
105 & MYL & Mylan N.V. & Health Care & HC \\ \hline
106 & SYK & Stryker Corp. & Health Care & HC \\ \hline
107 & THC & Tenet Healthcare Corp & Health Care & HC \\ \hline
108 & TMO & Thermo Fisher Scientific & Health Care & HC \\ \hline
109 & UNH & United Health Group Inc. & Health Care & HC \\ \hline
110 & VAR & Varian Medical Systems & Health Care & HC \\ \hline
111 & AVY & Avery Dennison Corp & Industrials & ID \\ \hline
112 & BA & Boeing Company & Industrials & ID \\ \hline
113 & CAT & Caterpillar Inc. & Industrials & ID \\ \hline
114 & CMI & Cummins Inc. & Industrials & ID \\ \hline
115 & CSX & CSX Corp. & Industrials & ID \\ \hline
116 & CTAS & Cintas Corporation & Industrials & ID \\ \hline
117 & DE & Deere \& Co. & Industrials & ID \\ \hline
118 & DHR & Danaher Corp. & Industrials & ID \\ \hline
119 & DNB & The Dun \& Bradstreet Corporation & Industrials & ID \\ \hline
120 & DOV & Dover Corp. & Industrials & ID \\ \hline
121 & EMR & Emerson Electric Company & Industrials & ID \\ \hline
122 & ETN & Eaton Corporation & Industrials & ID \\ \hline
123 & EXPD & Expeditors International & Industrials & ID \\ \hline
124 & FDX & FedEx Corporation & Industrials & ID \\ \hline
125 & FLS & Flowserve Corporation & Industrials & ID \\ \hline
126 & GD & General Dynamics & Industrials & ID \\ \hline
127 & GE & General Electric & Industrials & ID \\ \hline
128 & GLW & Corning Inc. & Industrials & ID \\ \hline
129 & GWW & Grainger (W.W.) Inc. & Industrials & ID \\ \hline
130 & HON & Honeywell Int'l Inc. & Industrials & ID \\ \hline
131 & IR & Ingersoll-Rand PLC & Industrials & ID \\ \hline
132 & ITW & Illinois Tool Works & Industrials & ID \\ \hline
133 & JEC & Jacobs Engineering Group & Industrials & ID \\ \hline
134 & LMT & Lockheed Martin Corp. & Industrials & ID \\ \hline
135 & LUV & Southwest Airlines & Industrials & ID \\ \hline
136 & MAS & Masco Corp. & Industrials & ID \\ \hline
137 & MMM & 3M Company & Industrials & ID \\ \hline
138 & ROK & Rockwell Automation Inc. & Industrials & ID \\ \hline
139 & RTN & Raytheon Co. & Industrials & ID \\ \hline
140 & TXT & Textron Inc. & Industrials & ID \\ \hline
141 & UNP & Union Pacific & Industrials & ID \\ \hline
142 & UTX & United Technologies & Industrials & ID \\ \hline
143 & AAPL & Apple Inc. & Information Technology & IT \\ \hline
144 & ADI & Analog Devices, Inc. & Information Technology & IT \\ \hline
145 & ADP & Automatic Data Processing & Information Technology & IT \\ \hline
146 & AMAT & Applied Materials Inc & Information Technology & IT \\ \hline
147 & AMD & Advanced Micro Devices Inc & Information Technology & IT \\ \hline
148 & CA & CA, Inc. & Information Technology & IT \\ \hline
149 & HPQ & HP Inc. & Information Technology & IT \\ \hline
150 & HRS & Harris Corporation & Information Technology & IT \\ \hline

\end{tabular}
\end{table}
\begin{table}[]
\centering
\begin{tabular}{|l|l|l|l|l|}
\hline
151 & IBM & International Business Machines & Information Technology & IT \\ \hline
152 & INTC & Intel Corp. & Information Technology & IT \\ \hline
153 & KLAC & KLA-Tencor Corp. & Information Technology & IT \\ \hline
154 & LRCX & Lam Research & Information Technology & IT \\ \hline
155 & MSI & Motorola Solutions Inc. & Information Technology & IT \\ \hline
156 & MU & Micron Technology & Information Technology & IT \\ \hline
157 & TSS & Total System Services, Inc. & Information Technology & IT \\ \hline
158 & TXN & Texas Instruments & Information Technology & IT \\ \hline
159 & WDC & Western Digital & Information Technology & IT \\ \hline
160 & XRX & Xerox Corp. & Information Technology & IT \\ \hline
161 & AA & Alcoa Corporation & Materials & MT \\ \hline
162 & APD & Air Products \& Chemicals Inc & Materials & MT \\ \hline
163 & BLL & Ball Corp & Materials & MT \\ \hline
164 & BMS & Bemis Company, Inc. & Materials & MT \\ \hline
165 & CLF & Cleveland-Cliffs Inc. & Materials & MT \\ \hline
166 & DD & DuPont & Materials & MT \\ \hline
167 & ECL & Ecolab Inc. & Materials & MT \\ \hline
168 & FMC & FMC Corporation & Materials & MT \\ \hline
169 & IFF & Intl Flavors \& Fragrances & Materials & MT \\ \hline
170 & IP & International Paper & Materials & MT \\ \hline
171 & NEM & Newmont Mining Corporation & Materials & MT \\ \hline
172 & PPG & PPG Industries & Materials & MT \\ \hline
173 & VMC & Vulcan Materials & Materials & MT \\ \hline
174 & CTL & CenturyLink Inc & Telecommunication Services & TC \\ \hline
175 & FTR & Frontier Communications Corporation & Telecommunication Services & TC \\ \hline
176 & S & Sprint Nextel Corp. & Telecommunication Services & TC \\ \hline
177 & T & AT\&T Inc & Telecommunication Services & TC \\ \hline
178 & VZ & Verizon Communications & Telecommunication Services & TC \\ \hline
179 & AEP & American Electric Power & Utilities & UT \\ \hline
180 & CMS & CMS Energy & Utilities & UT \\ \hline
181 & CNP & CenterPoint Energy & Utilities & UT \\ \hline
182 & D & Dominion Energy & Utilities & UT \\ \hline
183 & DTE & DTE Energy Co. & Utilities & UT \\ \hline
184 & ED & Consolidated Edison & Utilities & UT \\ \hline
185 & EIX & Edison Int'l & Utilities & UT \\ \hline
186 & EQT & EQT Corporation & Utilities & UT \\ \hline
187 & ETR & Entergy Corp. & Utilities & UT \\ \hline
188 & EXC & Exelon Corp. & Utilities & UT \\ \hline
189 & NEE & NextEra Energy & Utilities & UT \\ \hline
190 & NI & NiSource Inc. & Utilities & UT \\ \hline
191 & PNW & Pinnacle West Capital & Utilities & UT \\ \hline
192 & SO & Southern Co. & Utilities & UT \\ \hline
193 & WEC & Wec Energy Group Inc & Utilities & UT \\ \hline
194 & XEL & Xcel Energy Inc & Utilities & UT \\ \hline

\end{tabular}
\end{table}

\begin{table}[]
\centering
\caption{List of all stocks of Japan market (Nikkei 225) considered for the analysis. The first column has the serial number,
the second column has the abbreviation, the third column has the full name of the stock, and the
fourth column specifies the sector as given in the Nikkei 225.}
\label{JPN_Table}
\begin{tabular}{|l|l|l|l|l|}
\hline
\textbf{S.No.} & \textbf{Code} & \textbf{Company Name} & \textbf{Sector} & \textbf{Abbrv} \\ \hline
1 & S-8801 & MITSUI FUDOSAN CO., LTD. & Capital Goods & CG \\ \hline
2 & S-8802 & MITSUBISHI ESTATE CO., LTD. & Capital Goods & CG \\ \hline
3 & S-8804 & TOKYO TATEMONO CO., LTD. & Capital Goods & CG \\ \hline
4 & S-8830 & SUMITOMO REALTY \& DEVELOPMENT CO., LTD. & Capital Goods & CG \\ \hline
5 & S-7003 & MITSUI ENG. \& SHIPBUILD. CO., LTD. & Capital Goods & CG \\ \hline
6 & S-7012 & KAWASAKI HEAVY IND., LTD. & Capital Goods & CG \\ \hline
7 & S-9202 & ANA HOLDINGS INC. & Capital Goods & CG \\ \hline
8 & S-1801 & TAISEI CORP. & Capital Goods & CG \\ \hline
9 & S-1802 & OBAYASHI CORP. & Capital Goods & CG \\ \hline
10 & S-1803 & SHIMIZU CORP. & Capital Goods & CG \\ \hline
11 & S-1808 & HASEKO CORP. & Capital Goods & CG \\ \hline
12 & S-1812 & KAJIMA CORP. & Capital Goods & CG \\ \hline
13 & S-1925 & DAIWA HOUSE IND. CO., LTD. & Capital Goods & CG \\ \hline
14 & S-1928 & SEKISUI HOUSE, LTD. & Capital Goods & CG \\ \hline
15 & S-1963 & JGC CORP. & Capital Goods & CG \\ \hline
16 & S-5631 & THE JAPAN STEEL WORKS, LTD. & Capital Goods & CG \\ \hline
17 & S-6103 & OKUMA CORP. & Capital Goods & CG \\ \hline
18 & S-6113 & AMADA HOLDINGS CO., LTD. & Capital Goods & CG \\ \hline
19 & S-6301 & KOMATSU LTD. & Capital Goods & CG \\ \hline
20 & S-6302 & SUMITOMO HEAVY IND., LTD. & Capital Goods & CG \\ \hline
21 & S-6305 & HITACHI CONST. MACH. CO., LTD. & Capital Goods & CG \\ \hline
22 & S-6326 & KUBOTA CORP. & Capital Goods & CG \\ \hline
23 & S-6361 & EBARA CORP. & Capital Goods & CG \\ \hline
24 & S-6366 & CHIYODA CORP. & Capital Goods & CG \\ \hline
25 & S-6367 & DAIKIN INDUSTRIES, LTD. & Capital Goods & CG \\ \hline
26 & S-6471 & NSK LTD. & Capital Goods & CG \\ \hline
27 & S-6472 & NTN CORP. & Capital Goods & CG \\ \hline
28 & S-6473 & JTEKT CORP. & Capital Goods & CG \\ \hline
29 & S-7004 & HITACHI ZOSEN CORP. & Capital Goods & CG \\ \hline
30 & S-7011 & MITSUBISHI HEAVY IND., LTD. & Capital Goods & CG \\ \hline
31 & S-7013 & IHI CORP. & Capital Goods & CG \\ \hline
32 & S-7911 & TOPPAN PRINTING CO., LTD. & Capital Goods & CG \\ \hline
33 & S-7912 & DAI NIPPON PRINTING CO., LTD. & Capital Goods & CG \\ \hline
34 & S-7951 & YAMAHA CORP. & Capital Goods & CG \\ \hline
35 & S-1332 & NIPPON SUISAN KAISHA, LTD. & Consumer Goods & CN \\ \hline
36 & S-2002 & NISSHIN SEIFUN GROUP INC. & Consumer Goods & CN \\ \hline
37 & S-2282 & NH FOODS LTD. & Consumer Goods & CN \\ \hline
38 & S-2501 & SAPPORO HOLDINGS LTD. & Consumer Goods & CN \\ \hline
39 & S-2502 & ASAHI GROUP HOLDINGS, LTD. & Consumer Goods & CN \\ \hline
40 & S-2503 & KIRIN HOLDINGS CO., LTD. & Consumer Goods & CN \\ \hline
41 & S-2531 & TAKARA HOLDINGS INC. & Consumer Goods & CN \\ \hline
42 & S-2801 & KIKKOMAN CORP. & Consumer Goods & CN \\ \hline
43 & S-2802 & AJINOMOTO CO., INC. & Consumer Goods & CN \\ \hline
44 & S-2871 & NICHIREI CORP. & Consumer Goods & CN \\ \hline
45 & S-8233 & TAKASHIMAYA CO., LTD. & Consumer Goods & CN \\ \hline
46 & S-8252 & MARUI GROUP CO., LTD. & Consumer Goods & CN \\ \hline
47 & S-8267 & AEON CO., LTD. & Consumer Goods & CN \\ \hline
48 & S-9602 & TOHO CO., LTD & Consumer Goods & CN \\ \hline
49 & S-9681 & TOKYO DOME CORP. & Consumer Goods & CN \\ \hline
50 & S-9735 & SECOM CO., LTD. & Consumer Goods & CN \\ \hline

\end{tabular}
\end{table}

\begin{table}[]
\centering
\begin{tabular}{|l|l|l|l|l|}
\hline
51 & S-8331 & THE CHIBA BANK, LTD. & Financials & FN \\ \hline
52 & S-8355 & THE SHIZUOKA BANK, LTD. & Financials & FN \\ \hline
53 & S-8253 & CREDIT SAISON CO., LTD. & Financials & FN \\ \hline
54 & S-8601 & DAIWA SECURITIES GROUP INC. & Financials & FN \\ \hline
55 & S-8604 & NOMURA HOLDINGS, INC. & Financials & FN \\ \hline
56 & S-3405 & KURARAY CO., LTD. & Materials & MT \\ \hline
57 & S-3407 & ASAHI KASEI CORP. & Materials & MT \\ \hline
58 & S-4004 & SHOWA DENKO K.K. & Materials & MT \\ \hline
59 & S-4005 & SUMITOMO CHEMICAL CO., LTD. & Materials & MT \\ \hline
60 & S-4021 & NISSAN CHEMICAL IND., LTD. & Materials & MT \\ \hline
61 & S-4042 & TOSOH CORP. & Materials & MT \\ \hline
62 & S-4043 & TOKUYAMA CORP. & Materials & MT \\ \hline
63 & S-4061 & DENKA CO., LTD. & Materials & MT \\ \hline
64 & S-4063 & SHIN-ETSU CHEMICAL CO., LTD. & Materials & MT \\ \hline
65 & S-4183 & MITSUI CHEMICALS, INC. & Materials & MT \\ \hline
66 & S-4208 & UBE INDUSTRIES, LTD. & Materials & MT \\ \hline
67 & S-4272 & NIPPON KAYAKU CO., LTD. & Materials & MT \\ \hline
68 & S-4452 & KAO CORP. & Materials & MT \\ \hline
69 & S-4901 & FUJIFILM HOLDINGS CORP. & Materials & MT \\ \hline
70 & S-4911 & SHISEIDO CO., LTD. & Materials & MT \\ \hline
71 & S-6988 & NITTO DENKO CORP. & Materials & MT \\ \hline
72 & S-5002 & SHOWA SHELL SEKIYU K.K. & Materials & MT \\ \hline
73 & S-5201 & ASAHI GLASS CO., LTD. & Materials & MT \\ \hline
74 & S-5202 & NIPPON SHEET GLASS CO., LTD. & Materials & MT \\ \hline
75 & S-5214 & NIPPON ELECTRIC GLASS CO., LTD. & Materials & MT \\ \hline
76 & S-5232 & SUMITOMO OSAKA CEMENT CO., LTD. & Materials & MT \\ \hline
77 & S-5233 & TAIHEIYO CEMENT CORP. & Materials & MT \\ \hline
78 & S-5301 & TOKAI CARBON CO., LTD. & Materials & MT \\ \hline
79 & S-5332 & TOTO LTD. & Materials & MT \\ \hline
80 & S-5333 & NGK INSULATORS, LTD. & Materials & MT \\ \hline
81 & S-5706 & MITSUI MINING \& SMELTING CO. & Materials & MT \\ \hline
82 & S-5707 & TOHO ZINC CO., LTD. & Materials & MT \\ \hline
83 & S-5711 & MITSUBISHI MATERIALS CORP. & Materials & MT \\ \hline
84 & S-5713 & SUMITOMO METAL MINING CO., LTD. & Materials & MT \\ \hline
85 & S-5714 & DOWA HOLDINGS CO., LTD. & Materials & MT \\ \hline
86 & S-5715 & FURUKAWA CO., LTD. & Materials & MT \\ \hline
87 & S-5801 & FURUKAWA ELECTRIC CO., LTD. & Materials & MT \\ \hline
88 & S-5802 & SUMITOMO ELECTRIC IND., LTD. & Materials & MT \\ \hline
89 & S-5803 & FUJIKURA LTD. & Materials & MT \\ \hline
90 & S-5901 & TOYO SEIKAN GROUP HOLDINGS, LTD. & Materials & MT \\ \hline
91 & S-3865 & HOKUETSU KISHU PAPER CO., LTD. & Materials & MT \\ \hline
92 & S-3861 & OJI HOLDINGS CORP. & Materials & MT \\ \hline
93 & S-5101 & THE YOKOHAMA RUBBER CO., LTD. & Materials & MT \\ \hline
94 & S-5108 & BRIDGESTONE CORP. & Materials & MT \\ \hline
95 & S-5401 & NIPPON STEEL \& SUMITOMO METAL CORP. & Materials & MT \\ \hline
96 & S-5406 & KOBE STEEL, LTD. & Materials & MT \\ \hline
97 & S-5541 & PACIFIC METALS CO., LTD. & Materials & MT \\ \hline
98 & S-3101 & TOYOBO CO., LTD. & Materials & MT \\ \hline
99 & S-3103 & UNITIKA, LTD. & Materials & MT \\ \hline
100 & S-3401 & TEIJIN LTD. & Materials & MT \\ \hline

\end{tabular}
\end{table}

\begin{table}[]
\centering
\begin{tabular}{|l|l|l|l|l|}
\hline
101 & S-3402 & TORAY INDUSTRIES, INC. & Materials & MT \\ \hline
102 & S-8001 & ITOCHU CORP. & Materials & MT \\ \hline
103 & S-8002 & MARUBENI CORP. & Materials & MT \\ \hline
104 & S-8015 & TOYOTA TSUSHO CORP. & Materials & MT \\ \hline
105 & S-8031 & MITSUI \& CO., LTD. & Materials & MT \\ \hline
106 & S-8053 & SUMITOMO CORP. & Materials & MT \\ \hline
107 & S-8058 & MITSUBISHI CORP. & Materials & MT \\ \hline
108 & S-4151 & KYOWA HAKKO KIRIN CO., LTD. & Pharmaceuticals & PH \\ \hline
109 & S-4503 & ASTELLAS PHARMA INC. & Pharmaceuticals & PH \\ \hline
110 & S-4506 & SUMITOMO DAINIPPON PHARMA CO., LTD. & Pharmaceuticals & PH \\ \hline
111 & S-4507 & SHIONOGI \& CO., LTD. & Pharmaceuticals & PH \\ \hline
112 & S-4519 & CHUGAI PHARMACEUTICAL CO., LTD. & Pharmaceuticals & PH \\ \hline
113 & S-4523 & EISAI CO., LTD. & Pharmaceuticals & PH \\ \hline
114 & S-7201 & NISSAN MOTOR CO., LTD. & Information Technology & IT \\ \hline
115 & S-7202 & ISUZU MOTORS LTD. & Information Technology & IT \\ \hline
116 & S-7205 & HINO MOTORS, LTD. & Information Technology & IT \\ \hline
117 & S-7261 & MAZDA MOTOR CORP. & Information Technology & IT \\ \hline
118 & S-7267 & HONDA MOTOR CO., LTD. & Information Technology & IT \\ \hline
119 & S-7270 & SUBARU CORP. & Information Technology & IT \\ \hline
120 & S-7272 & YAMAHA MOTOR CO., LTD. & Information Technology & IT \\ \hline
121 & S-3105 & NISSHINBO HOLDINGS INC. & Information Technology & IT \\ \hline
122 & S-6479 & MINEBEA MITSUMI INC. & Information Technology & IT \\ \hline
123 & S-6501 & HITACHI, LTD. & Information Technology & IT \\ \hline
124 & S-6502 & TOSHIBA CORP. & Information Technology & IT \\ \hline
125 & S-6503 & MITSUBISHI ELECTRIC CORP. & Information Technology & IT \\ \hline
126 & S-6504 & FUJI ELECTRIC CO., LTD. & Information Technology & IT \\ \hline
127 & S-6506 & YASKAWA ELECTRIC CORP. & Information Technology & IT \\ \hline
128 & S-6508 & MEIDENSHA CORP. & Information Technology & IT \\ \hline
129 & S-6701 & NEC CORP. & Information Technology & IT \\ \hline
130 & S-6702 & FUJITSU LTD. & Information Technology & IT \\ \hline
131 & S-6703 & OKI ELECTRIC IND. CO., LTD. & Information Technology & IT \\ \hline
132 & S-6752 & PANASONIC CORP. & Information Technology & IT \\ \hline
133 & S-6758 & SONY CORP. & Information Technology & IT \\ \hline
134 & S-6762 & TDK CORP. & Information Technology & IT \\ \hline
135 & S-6770 & ALPS ELECTRIC CO., LTD. & Information Technology & IT \\ \hline
136 & S-6773 & PIONEER CORP. & Information Technology & IT \\ \hline
137 & S-6841 & YOKOGAWA ELECTRIC CORP. & Information Technology & IT \\ \hline
138 & S-6902 & DENSO CORP. & Information Technology & IT \\ \hline
139 & S-6952 & CASIO COMPUTER CO., LTD. & Information Technology & IT \\ \hline
140 & S-6954 & FANUC CORP. & Information Technology & IT \\ \hline
141 & S-6971 & KYOCERA CORP. & Information Technology & IT \\ \hline
142 & S-6976 & TAIYO YUDEN CO., LTD. & Information Technology & IT \\ \hline
143 & S-7752 & RICOH CO., LTD. & Information Technology & IT \\ \hline
144 & S-8035 & TOKYO ELECTRON LTD. & Information Technology & IT \\ \hline
145 & S-4543 & TERUMO CORP. & Information Technology & IT \\ \hline
146 & S-4902 & KONICA MINOLTA, INC. & Information Technology & IT \\ \hline
147 & S-7731 & NIKON CORP. & Information Technology & IT \\ \hline
148 & S-7733 & OLYMPUS CORP. & Information Technology & IT \\ \hline
149 & S-7762 & CITIZEN WATCH CO., LTD. & Information Technology & IT \\ \hline
150 & S-9501 & TOKYO ELECTRIC POWER COMPANY HOLDINGS, I & Transportation \& Utilities & TU \\ \hline

\end{tabular}
\end{table}

\begin{table}[]
\centering
\begin{tabular}{|l|l|l|l|l|}
\hline
151 & S-9502 & CHUBU ELECTRIC POWER CO., INC. & Transportation \& Utilities & TU \\ \hline
152 & S-9503 & THE KANSAI ELECTRIC POWER CO., INC. & Transportation \& Utilities & TU \\ \hline
153 & S-9531 & TOKYO GAS CO., LTD. & Transportation \& Utilities & TU \\ \hline
154 & S-9532 & OSAKA GAS CO., LTD. & Transportation \& Utilities & TU \\ \hline
155 & S-9062 & NIPPON EXPRESS CO., LTD. & Transportation \& Utilities & TU \\ \hline
156 & S-9064 & YAMATO HOLDINGS CO., LTD. & Transportation \& Utilities & TU \\ \hline
157 & S-9101 & NIPPON YUSEN K.K. & Transportation \& Utilities & TU \\ \hline
158 & S-9104 & MITSUI O.S.K.LINES, LTD. & Transportation \& Utilities & TU \\ \hline
159 & S-9107 & KAWASAKI KISEN KAISHA, LTD. & Transportation \& Utilities & TU \\ \hline
160 & S-9001 & TOBU RAILWAY CO., LTD. & Transportation \& Utilities & TU \\ \hline
161 & S-9005 & TOKYU CORP. & Transportation \& Utilities & TU \\ \hline
162 & S-9007 & ODAKYU ELECTRIC RAILWAY CO., LTD. & Transportation \& Utilities & TU \\ \hline
163 & S-9008 & KEIO CORP. & Transportation \& Utilities & TU \\ \hline
164 & S-9009 & KEISEI ELECTRIC RAILWAY CO., LTD. & Transportation \& Utilities & TU \\ \hline
165 & S-9301 & MITSUBISHI LOGISTICS CORP. & Transportation \& Utilities & TU \\ \hline

\end{tabular}
\end{table}



\end{document}